\documentclass[revtex4]{emulateapj}

\newcommand{\Teff}{\mbox{$T_{\rm eff}$}}
\newcommand{\logg}{\mbox{$\log g$}}
\usepackage{float}
\usepackage{lscape}
\usepackage{url}
\usepackage{longtable}
\usepackage{enumerate}

\usepackage[normalem]{ulem}

\usepackage{color}

\usepackage{multirow}
\usepackage{graphicx}
\usepackage{rotating}
\usepackage{footnote}

\usepackage{amsmath}
\usepackage{amssymb}
\usepackage{mathrsfs}

\makeatletter

\newcommand{\Rmnum}[1]{\expandafter\@slowromancap\romannumeral #1@}
\makeatother

\begin{document}

%%\title{EXTRASOLAR STORMS: PHASE SHIFTS IN SIMULTANEOUS \emph{HUBBLE} AND \emph{SPITZER} LIGHT CURVES OF BROWN DWARFS}
%\title{EXTRASOLAR STORMS: PRESSURE-DEPENDENT CHANGES IN LIGHT CURVE \sout{\red{SHAPE AND}} PHASE IN BROWN DWARFS 
%       FROM SIMULTANEOUS \emph{HUBBLE} AND \emph{SPITZER} OBSERVATIONS}
\title{\emph{EXTRASOLAR STORMS}: PRESSURE-DEPENDENT CHANGES IN LIGHT CURVE PHASE IN BROWN DWARFS 
       FROM SIMULTANEOUS \emph{HUBBLE} AND \emph{SPITZER} OBSERVATIONS}

\author{Hao Yang}
\affil{Department of Astronomy, University of Arizona, 933 N. Cherry Avenue, Tucson, AZ 85721, USA} 
\email{haoyang@email.arizona.edu}

\author{D\'aniel Apai}
\affil{Department of Astronomy, University of Arizona, 933 N. Cherry Avenue, Tucson, AZ 85721, USA;} 
\affil{and Department of Planetary Sciences, 1629 E. University Blvd, Tucson, AZ 85721, USA;}
\affil{Earths in Other Solar Systems Team}
\email{apai@arizona.edu}

\author{Mark S. Marley}
\affil{NASA Ames Research Center, Naval Air Station, Moffett Field, Mountain View, CA 94035, USA} 
\author{Theodora Karalidi}
\affil{Department of Astronomy, University of Arizona, 933 N. Cherry Avenue, Tucson, AZ 85721, USA} 
\author{Davin Flateau}
\affil{Department of Planetary Sciences, 1629 E. University Blvd, Tucson, AZ 85721, USA}
\author{Adam P. Showman}
\affil{Department of Planetary Sciences, University of Arizona, 1629 University Boulevard, Tucson, AZ 85721, USA} 
\author{Stanimir Metchev}
\affil{The University of Western Ontario, Department of Physics \& Astronomy, Centre for Planetary Science and Exploration, 1151 Richmond St, London, ON N6A 3K7, Canada}
\affil{and Stony Brook University, Department of Physics \& Astronomy, 100 Nicolls Rd, Stony Brook, NY 11794-3800, USA}
\author{Esther Buenzli}
\affil{Institute for Astronomy, ETH Z\"urich Wolfgang-Pauli-Str. 27, 8093 Z\"urich, Switzerland}
\author{Jacqueline Radigan}
\affil{Space Telescope Science Institute, 3700 San Martin Drive,Baltimore, MD 21218, USA} 
\author{\'Etienne Artigau}
\affil{D\'epartement de Physique, Universit\'e de Montr\'eal, C.P. 6128 Succ. Centre-ville, Montr\'eal, QC H3C 3J7, Canada} 
\author{Patrick J. Lowrance}
\affil{Infrared Processing and Analysis Center, MS 100-22, California Institute of Technology, Pasadena, CA 91125, USA} 
\author{Adam J. Burgasser}
\affil{Center for Astrophysics and Space Science, University of California San Diego, La Jolla, CA 92093, USA}
%%

%%abstract
\begin{abstract}

We present \emph{Spitzer}/IRAC Ch1 and Ch2 monitoring of six brown dwarfs during 8 different epochs over the course of 20 months. 
For four {brown} dwarfs, we also obtained simulataneous \emph{HST}/WFC3 G141 Grism spectra during two epochs and derived light curves 
in five narrow-band filters. 
Probing different pressure levels in the atmospheres, the multi-wavelength light curves of our six targets all exhibit variations, 
and the shape of the light curves evolves over the timescale of a rotation period, ranging from 1.4 h to 13 h. 
We compare the shapes of the light curves and estimate the phase shifts between the light curves observed at different wavelengths 
by comparing the phase of the primary Fourier components. We use state-of-the-art atmosphere models to determine the flux contribution 
of different pressure layers to the observed flux in each filter.
We find that the light curves that probe higher pressures are similar and in phase, but are offset and often different from 
the light curves that probe lower pressures. The phase differences between the two groups of light curves suggest that the modulations 
seen at lower and higher pressures may be introduced by different cloud layers.

\end{abstract}


\section{INTRODUCTION}

%OUTLINE:
%%- Current theoretical understanding of the BD variabilities. Patchy cloud models, temperature perturbation model, circulation models.
{Sharing the same range of effective temperatures with many extrasolar giant planets, brown dwarfs are thought to possess similar atmospheric properties 
\citep[e.g.,][]{lodders2006}. Compared with planets, brown dwarfs are generally more luminous and have no disburbance from host stars. Therefore, 
studies of brown dwarf atmospheres can be performed with higher precision, providing important reference and guidance to understanding of the atmospheres of extrasolar planets. }

The presence of condensate clouds in brown dwarf atmospheres strongly impacts the chemistry and thermal structure in the atmosphere, 
removing gas-phase opacity and modifying the emergent spectra \citep[e.g.,][]{ackerman2001, cooper2003, helling2008, marley2015}.
Cloud covers are likely to be heterogenous in these ultracool atmospheres, 
similar to what have been observed for solar system giant planets \citep[e.g.,][]{west2009, sromovsky2012, karalidi2015, simon2015},
and complex atmospheric structures such as spots and storms combined with rotational modulation are expected to 
produce periodic photometric variabilities \citep{bailerjones2001}.
In addition to patch clouds, {  
large-scale atmospheric motions and temperature fluctuations, induced by regional and global atmospheric circulation,
have also been proposed to explain the brightness variations \citep{showman2013, zhang2014, robinson2014}.  }

%2.recent BD variability studies
{Starting in the} early 2000s, substantial observational efforts have been put into detecting and characterizing periodic flux variations of brown dwarfs.
While most early ground-based monitoring campaigns {had} no or marginal detections \citep[e.g.,][]{bailerjones2001, clarke2002, koen2005, clarke2008, khandrika2013}, 
\citet{artigau2009} found that the L/T transition dwarf, SIMP J013656.5+093347 (hereafter SIMP0136), displayed periodic variation in the $J$-band 
with an amplitude of $50$ mmag, and \citet{radigan2012} measured a peak-to-peak amplitude as large as 26\% in the $J$-band light curves of another L/T transition
object, 2MASS J21392676+0220226 (hereafter 2M2139). In a large ground-based $J$-band survey of 57 brown dwarfs, \citet{radigan2014a} concluded that 
strong flux variations over 2\% were exclusively observed for the L/T transition objects in their sample.
With much higher precision than ground-based observations, 
space-based photometric and spectroscopic monitoring campaigns have revealed that low-amplitude flux variations 
on a sub-percent level appear to be a common characteristic for all L and T dwarfs \citep{heinze2013,buenzli2014, metchev2015}.  

While single-band photometric monitoring uncovers the longitudinal cloud structures in the brown dwarfs,
multi-wavelength observations probe different depths in the brown dwarf atmospheres and provide valuable information on the vertical cloud structures. 
So far there have been only a handful of such studies.
\citet{buenzli2012} discovered for the late-T dwarf 2MASS J22282889-4310262 (hereafter 2M2228) 
that simultaneous \emph{HST} narrow-band and \emph{Spitzer} 4.5-$\mu$m light curves 
have different phases. They derived from models the pressure levels probed by different wavelengths and found the phase shifts increase with decreasing
pressure level, indicating complex atmospheric structures in both horizontal and vertical directions. 
\citet{biller2013} performed ground-based photometric monitoring of the T0.5 dwarf Luhman 16B \citep{luhman2013, gillon2013}
 and their simultaneous multi-band light curves suggested 
similar pressure-dependent phase shifts. On the other hand, in the \emph{HST} observations of Luhman 16B \citet{buenzli2015a} found no phase shifts between
light curves of narrow $J$ and $H$ bands as well as the 1.35-1.44 $\mu$m water band. This is consistent with the results reported by \citet{apai2013} 
that the same \emph{HST} narrow-band light curves showed no phase shifts for the two L/T transition objects, SIMP0136 and 2M2139.
 \citet{gizis2015} found that the \emph{Kepler} optical light curves of the L1 dwarf WISEP J190648.47+401106.8 are consistent 
in phase with subsequent \emph{Spitzer} 4.5-$\mu$m light curves obtained about 5 months after the \emph{Kepler} monitoring ended.
To date, 2M2228 remains the only brown dwarf found to display unambiguous phase shifts between multi-band light curves, and {observationally
it is difficult to probe the cloud structures in the vertical direction.}

%- No good explanation of the observed phase shifts in 2M2228.
%%To date, there have been only a few studies on the vertical atmospheric structures of brown dwarfs.

In this paper, we report the first results from the \emph{Spitzer} Cycle-9 Exploration Science Program, \emph{Extrasolar Storms}, 
on detecting phase shifts between the light curves of six varying brown dwarfs observed in two IRAC channels 
as well as between simultaneous \emph{HST} and \emph{Spitzer} light curves of four objects. 
The rest of this paper is organized as follows. In \S 2, we give the background information on the the six targets in our sample.
In \S 3, we describe our observations and the data reduction procedures. In \S 4, we discuss our methods for measuring the phase shifts
in our \emph{Spitzer} and \emph{HST} observations. The phase shift measurements and other results are presented in \S 5, which is followed by some discussion in \S 6.

\section{TARGETS}

%All six targets are known to variable. 

\subsection{2MASS J15074769-1627386 and 2MASS J18212815+1414010 (L5)}

The L5 dwarf 2MASS J15074769-1627386 (hereafter 2M1507) was discovered by \citet{reid2000}
and the L5 dwarf 2MASS J18212815+1414010 (hereafter 2M1821) was discovered by \citet{looper2008}.
Spectroscopic analysis by \citet{gagne2014} identified 2M1821 as a young field dwarf showing signs of low surface gravity.
\citet{sahlmann2015} measured the parallax of 2M1821 to be $106.15 \pm 0.18$ mas, corresponding to a distance of $9.38 \pm 0.03$ pc.
Based on the $J-K$ color of 1.78, weak water bands, triangular $H$-band continuum, and strong 9-11 $\mu$m silicate absorption, 
\citet{looper2008} suggested that the atmosphere of 2M1821 might have unusually thick dust clouds.
For 2M1507,  weak silicate absorption between 9 and 11 $\mu$m was found \citet{cushing2006}, indicating different cloud thickness than 2M1821.

With the \emph{HST}/WFC3 G141 grism spectra of 2M1507 and 2M1821,  \citet{yang2015} studied the wavelength dependence of 
their flux density variation between 1.1 and 1.7 $\mu$m. By taking the ratio of the brightest and faintest spectra in an \emph{HST}
visit, \citet{yang2015} found that for the two L5 dwarfs the flux density of the 1.4 $\mu$m water band varies at a similar rate
as that of the $J$- and $H$-band. This is significantly different from the reduced variation amplitude in the 1.4 $\mu$m water band 
observed for the two T2 dwarfs (SIMP J013656.5+093347.3 and 2MASS J21392676+0220226) reported by \citet{apai2013}. 
\citet{yang2015} showed that models of an L5 dwarf with a haze layer high ($<$ 50 mbar) in the atmosphere could explain 
the similar flux variation amplitude in and out of the 1.4 $\mu$m water band observed for 2M1507 and 2M1821.

\subsection{SIMP J013656.5+093347.3 and 2MASS J21392676+0220226 (T2)}

The T2 dwarf SIMP0136 was discovered by \citet{artigau2006} in a proper-motion survey.
At an estimated photometric distance of 6.4 $\pm$ 0.3 pc, it is the brightest T dwarf in the northern hemisphere.
Ground-based photometric monitoring by \citet{artigau2009} revealed that SIMP0136 is variable with a peak-to-peak amplitude 
of $\sim$ 50 mmag in $J$-band, marking the first detection of high-amplitude IR flux variations for a T dwarf.
\citet{kao2015} detected circularly polarized radio emission with two pulses in the 4-8 GHz band and estimated a magnetic 
field strength of at least 2.5 kG.

The T2 dwarf 2M2139 was discovered by \citet{reid2008}. 
\cite{radigan2012} reported periodic flux variation of 2M2139 in the $J-$band with a peak-to-peak amplitude as large as 26\%, 
making 2M2139 the most variable brown dwarf discovered to date. 

\emph{HST} time-resolved spectroscopy by \citet{apai2013} provided high-quality spectral series and light curves for both SIMP0136 and 2M2139. 
Based on these data and comparison to state-of-the-art atmosphere models the authors concluded that the variations in both T2 dwarfs
are caused by thickness variations in the silicate cloud cover (warm thin and cooler thicker clouds). 
Light curve modeling through a Genetic Algorithm-optimized ray tracer by \citet{apai2013} and an MCMC-optimized pixelized atmosphere 
model by \citet{karalidi2015} both found that at least three elliptical features are required to fit the light curve of SIMP0136 and 2M2139.
Principal components analysis of the HST spectral series showed that $>$97\% of the spectral variations can be reproduced with 
only a single principal component on top of a mean spectrum, arguing for a single type of cloud feature, 
consistent with the result of the atmospheric modeling.

\subsection{2MASS J13243553+6358281 (T2)}
The T2 dwarf 2MASS J13243553+6358281 (hereafter 2M1324) was discovered independently by \citet{looper2007} and \citet{metchev2008}.
It has an usually red spectral energy distribution and also exhibits peculiar IR colors \citep{looper2007, metchev2008, faherty2009}.
\citet{looper2007}, \citet{burgasser2010a}, and \citet{geibler2011} discussed the possibility of 2M1324 being a close binary
(L9 + T2 or L8 + T3.5).

\subsection{2MASS J22282889-4310262 (T6)}
The T6 dwarf 2M2228 was discovered by \citet{burgasser2003}.
In a ground-based monitoring campaign searching for variability, \citet{clarke2008} detected a peak-to-peak amplitude
of $15.4 \pm 1.4$ mmag for 2M2228. 

\citet{buenzli2012} analyzed \emph{HST}/WFC3 G141 grism spectra and partially simulatenous \emph{Spitzer} Ch2 photometry of 2M2228,
and for the first time found phase shifts between light curves of five narrow spectral bands between 1.1 and 1.7 $\mu$m and 4.5 $\mu$m. 
The fluxes in these spectral bands probe different depths in the atmosphere, and
the measured phase shifts of 2M2228 were found to correlate with the characteristic pressure levels derived from atmospheric models.
For lower pressure levels, the phase shift is larger. Such findings by \citet{buenzli2012} revealed atmospheric structures in the 
vertical direction, adding a new dimension in the studies of the ultracool atmospheres.

\section{OBSERVATIONS AND DATA REDUCTION}

With the \emph{Spitzer} Cycle-9 Exploration Science Program, \emph{Extrasolar Storms} (hereafter \emph{Storms}; Program ID: 90063, PI: D. Apai),
we obtained for six targets 1,144 hours of photometric data in staring mode in channel 1 (Ch1, the $3.6~\mu$m bandpass) 
and channel 2 (Ch2, the $4.5~\mu$m bandpass) of the \emph{Spitzer} Infrared Array Camera \citep[IRAC;][]{fazio2004}.
Our observations were designed to monitor the light curve evolution of each target over more than 1,000 rotation periods in eight separate 
\emph{Spitzer} visits, probing flux variations on a number of timescales and studying the evolution and dynamics of the brown dwarf 
atmospheres and their heterogenous cloud covers.
The target properties are provided in Table~\ref{obs1} and details of the observations are given in Table~\ref{obs2}.
Previously, \citet{metchev2015} have observed all the \emph{Storms} targets except for SIMP0136 for about 20 hr each. 

%2M1324 has also been observed with \emph{Spitzer} for 21 hr \citep{metchev2015}.
%2M1324 has also been observed with \emph{Spitzer} for 21 hr \citep{metchev2015}.
% have observed both 2M1507 and 2M1821 with\emph{Spitzer} for 20 and 16 hr, respectively. 

In addition to the \emph{Spitzer} observations, for the four shorter-period objects in the sample, we also 
obtained time-resolved, high-precision \emph{HST}/WFC3 G141 grism spectra simulateneously during two of the eight \emph{Spitzer} visits. 
As the coordinated \emph{HST} component (Program ID: 13176, PI: D. Apai) of the \emph{Storms} program, 
the WFC3 observations were acquired between April, 2013 and October, 2013 for a total of 28 orbits. 
Detailed information of the \emph{HST} observations are also listed in Table~\ref{obs1}.
The data reduction procedures are described in detail in \citet{apai2013} and \citet{yang2015}.
The reduced G141 grism spectra provide wavelength coverage between 1.05 and 1.7 $\mu$m and a spectral resolution of $\sim$ 130.
The signal-to-noise ratio is over 300. Part of the \emph{HST} observations have been published in \citet{yang2015}. 

\begin{figure*}[!bht]
  \begin{center}
    \includegraphics[angle=270, scale=0.7]{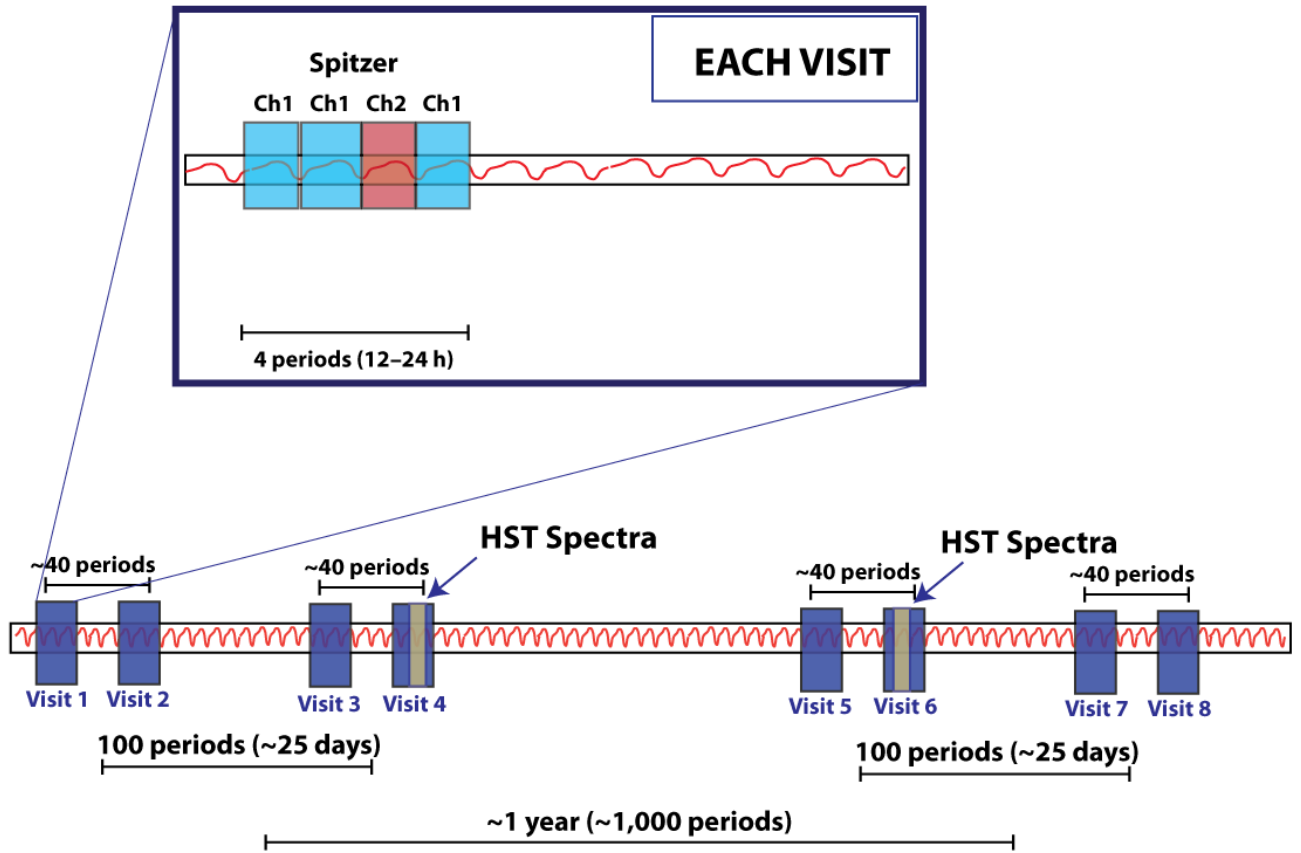}
           \label{fig1schedule}
              \end{center}
       \caption{Schematic illustration of how observations for the \emph{Extrasolar Storms} program are scheduled.}
    \end{figure*}

\begin{figure*}[ht]
  \begin{center}
    \includegraphics[scale=0.80]{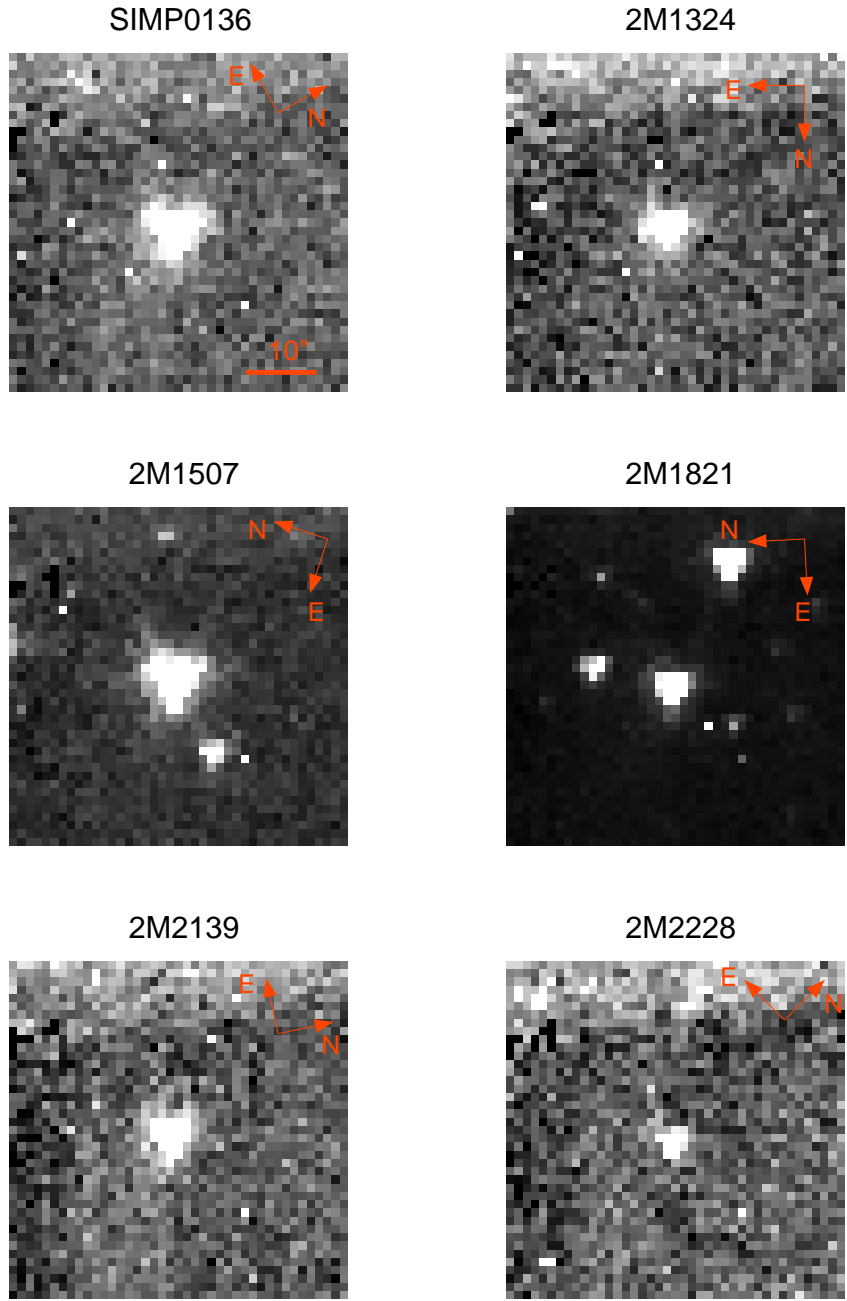}
       \caption{Sample \emph{Spitzer} Ch1 images of all six \emph{Storms} targets. Each panel shows a subarray (40 $\times 40$ pixels)
    of the full Ch1 detector array (256 $\times$ 256 pixels), corresponding to a field of view of approximately $48\arcsec \times 48 \arcsec$ in the sky. }
           \label{rawimage}
              \end{center}
    \end{figure*}

The \emph{Spitzer} observations were carried out from December, 2012 to August, 2014. 
The exposure time for each individual image is 10.40 s. The detector arrays in both IRAC channels are 256 $\times$ 256 pixels in size 
and the pixel size is $1\farcs2 \times 1\farcs2$, providing a field of view of $5\farcm2 \times 5\farcm2$. 
The eight visits for each target were arranged in pairs of two. The two visits within a pair
were separated by $\sim$ 40 rotation periods, and each pair was separated by $\sim$ 100 rotation periods. 
The first four visits and the last four visits were separated by about 1,000 rotation periods, corresponding to roughly one year.
As illustrated in Figure~\ref{fig1schedule}, such scheduling of the observations allowed us to detect flux variations  
on a broad range of timescales and to probe different physical processes reponsible for these changes.
During each visit, we first observed two consecutive rotations in Ch1, then one rotation in Ch2. 
followed by another rotation in Ch1, allowing us to separate light curve evolution in time from 
the wavelength dependency of the light curve. 
Figure~\ref{rawimage} shows a sample of raw \emph{Spitzer} images for the \emph{Storms} targets.

To reduce the \emph{Spitzer} photometric data we first downloaded from the \emph{Spitzer} Science Center the corrected Basic Calibrated Data (cBCD) images, 
which have been processed through the IRAC calibration pipeline (version: S19.1.0). 
After flat-fielding and manually masking out bad pixels and bright objects within a 20-pixel radius of the target, 
we used the IDL routine $box\_centroider$ supplied by the \emph{Spitzer} Science Center to measure the exact location of the object on the detector.
After subtracting the background level determined from an annulus between 12 and 20 pixels 
from the centroid position of the target, we performed photometry for each image using the IDL routine $aper$ with a fixed aperture of 2 pixels. 
%To calculate the sky background, we use an annulus around the target with an inner radius of 12 pixels and an outer radius of 20 pixels.
Then we rejected photometric points with centroid positions outside of 5$\sigma$ in $x$ or $y$ from a 25-point median-smoothed values.
Flux measurements that are outside of 3$\sigma$ of the 25-point median-smoothed light curve were also rejected. 
In each visit, less than 3\% of the photometric points were rejected.

The reduced photometric data displays the intra-pixel sensitivity variation in IRAC \citep{reach2005}, commonly referred to as the pixel phase effect, 
as the detector sensitivity varies slightly depending on where the exact location (on a sub-pixel scale) of the point source is. 
This effect manifests itself as flux discontinuities due to re-acquisition of targets between consecutive \emph{Spitzer} Astronomical Observation 
Requests (AORs) as well as zigzag-shaped flux variations within an AOR with a period around 40 minutes.

To correct for the pixel-phase effect, we model the observed light curves as a combination of astrophysical variations and sensitivity variations,
and simultaneously fit both model components with Markov Chain Monte Carlo (MCMC) simulations. The best-fit model for sensitivity variations 
is then removed from the observed light curves. 

We model the sensitivity variations as a quadratic function of the target's centroid location, ($x$,$y$),
in pixels (e.g., Knutson et al. 2008; Heinze et al. 2013): 
$$ Q(x,y)= 1 + p_1 \times x + p_2 \times y+ p_3 \times x^2 + p_4 \times y^2 + p_5 \times x \times y
                 \ , \,\,\,\,\,\, \eqno(1)$$

To account for simultaneous astrophysical variations of our targets, %% in the normalized light curve, 
we fit a third-order Fourier series to the normalized light curve:
%%$$ F(t) = 1 + A_i cos(2{\pi}t \div P/i + B_i sin(2{\pi}t \div P/i)     \ , \,\,\,\,\,\, \eqno(2)$$
$$\mathscr{F}(t) = 1 + \sum\limits_{i=1}^3 \left( A_i~\textrm{cos}~(\frac{2{\pi}i}{P}~t) + B_i~\textrm{sin}~(\frac{2{\pi}i}{P}~t)   \right) \ , \,\,\,\,\,\, \eqno(2)$$
%%%old equation: $$\mathscr{F}(t) = 1 + \sum\limits_{i=1}^3 {\Big(} A_i~\textrm{cos}~(\frac{2{\pi}t}{P/i}) + B_i~\textrm{sin}~(\frac{2{\pi}t}{P/i})  {\Big)} \ , \,\,\,\,\,\, \eqno(2)$$
where $t$ is the observation timestamp and $P$ is the rotation period of the object, which is fixed to the value in Table~\ref{obs1}. 

\begin{figure*}[ht]
  \begin{center}
    \includegraphics[scale=0.80]{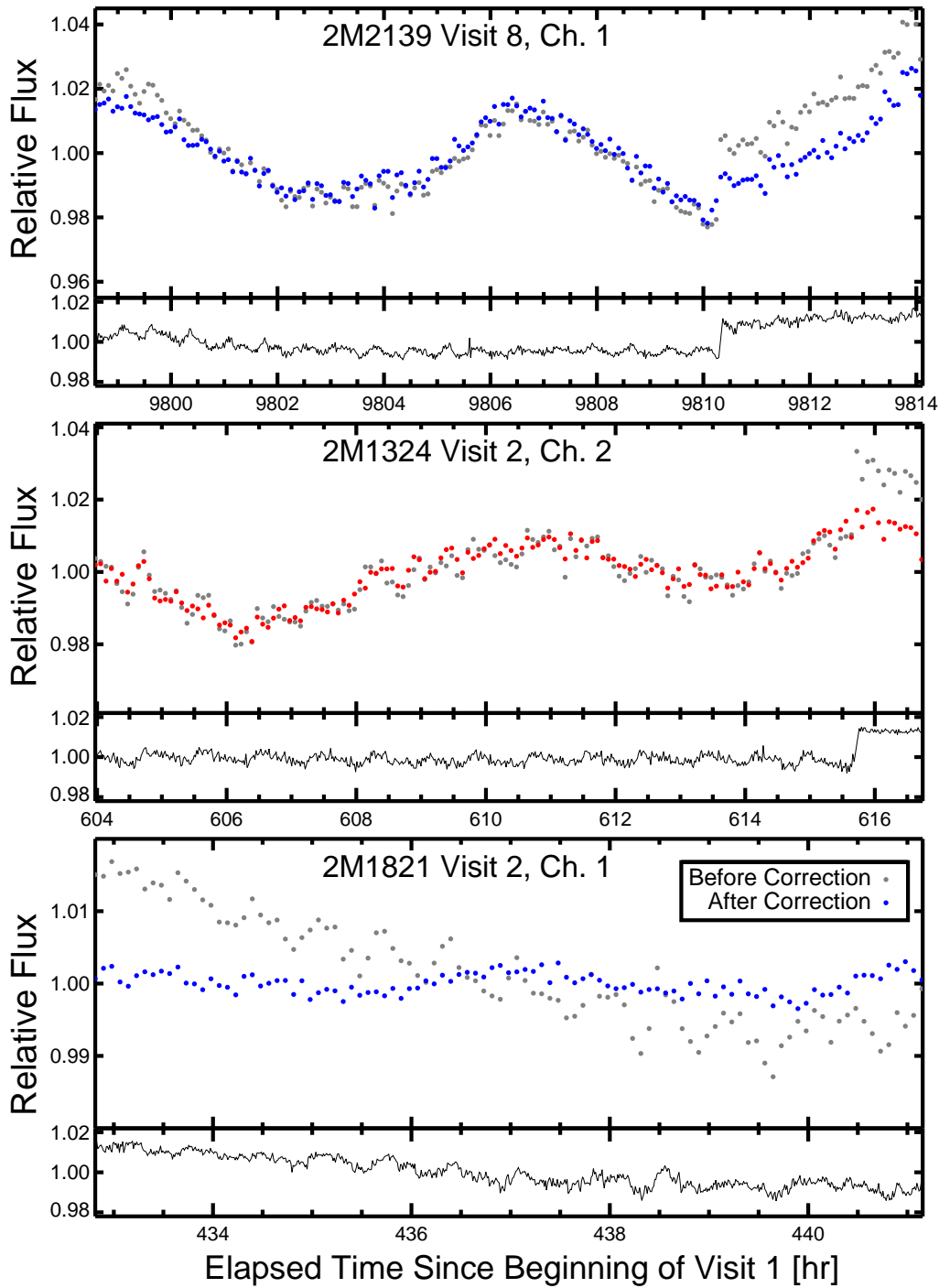}
       \vspace*{6mm}
       \caption{Three example light curves showing the correction for the intra-pixel sensitivity variations in the \emph{Spitzer} observations. 
 In each major panels, the gray dots are the extracted raw photometric points, while the red and blue dots show
 the corrected data after dividing out the best-fit quadratic correction functions. The corrections are able to remove sudden flux discontinuities 
 and zigzag-shaped flux changes caused by the intra-pixel sensitivity variations. 
 The minor panels display the best-fit quadractic correction functions for the specific visits.
 Note the different flux scales for the data and the quadractic correction function. The light curves are binned in 5-min intervals.}
           \label{mcmcexample}
              \end{center}
    \end{figure*}

Then we perform MCMC simulations utilizing the Python package PyMC \citep{patil2008}, 
and simulateneously fit the normalized observations with $Q(x,y)\mathscr{F}(t)$, the product of the quadractic function and the Fourier series.
The data in each channel of a visit are fitted separately, and for each run, two million iterations are calculated with additional $0.6$ million burn-in 
iterations. We have run five chains for multiple visits and found that different chains converge well within one million iterations and return the same 
results for the parameters. 
Finally, we corrected for the pixel phase effect by calculating the value of the best-fit quadratic correction function at each observation timestamp and 
dividing that value from the corresponding aperture photometric measurement. 
A few example light curves and the corresponding correction functions are shown in Figure~\ref{mcmcexample}. This correction method is able to
effectively remove the flux discontinuities due to target re-acquisition and flux oscillations due to target sub-pixel centroid shift.
For our subsequent analysis, all the corrected light curves were binned in 5-minute
intervals to reduce noise, and the midpoint of each 5-minute interval was calculated to be the time of observation.
Each of the three segments in a visit was normalized to its own mean value.

%SIMP0136: 2MASS J01365662+0933473; SIMP J013656.5+093347.3

\section{PHASE SHIFT ANALYSIS}

\subsection{Light Curve Shapes and Evolution}
%%1. Light curve evolution;
{ We found that the light curves in both \emph{Spitzer}/IRAC channels exhibit a variety of shapes, as shown in Figure~\ref{fig2replc}.
In some cases, the light curve shapes could evolve over a timescale as short as one rotation period, and the shape and amplitude of the light curve 
observed at different wavelengths can be substantially different.
}

{While our observations reveal differences in the light curve shape as a function of time and, sometimes, 
as a function of the wavelength of the observations,
in 20 out of 48 \emph{Spitzer} visits the differences between the light curves of consecutive rotations are relatively small. 
In these cases, the light curve shape observed at different wavelengths is either identical or very similar, 
but potentially delayed (``phase-shifted'') and may have different amplitudes.}

{In this paper, we performed analyses to quantify phase shifts of only the datasets where the light curves are slowly evolving and 
the features are similar between different wavelengths, so that simple, robust approaches can be applied to measure phase shifts. 
%%Group B contains datasets where the lightcurve shape is dissimilar enough between the different wavelengths that our simple phase shift measurements 
%%cannot be considered reliable. Table XX summarizes the datasets in the two groups.
%%Most of our datasets are in Group A and the rest of the paper focuses on their analysis.
A separate publication (Apai et al., in preparation) will fully explore the evolution of the light curves.}

{Here, we focus on the information available 
in the wavelength-dependent phase shifts. As different wavelengths of observation probe different pressure levels in the atmosphere, 
the phase shifts measured here can be used to explore the vertical-longitudinal structure of the atmosphere. 
This approach has been demonstrated in \citet{buenzli2012} and \citet{apai2013} observationally, and models by 
\citet{zhang2014} and \citet{robinson2014} have predicted characteristic observational signatures contained in such pressure-dependent phase shifts.  }

With the \emph{Storms} data set, we adopted different approaches to determine phase shifts for two different data subsets: 
non-simultaneous \emph{Spitzer} Ch1 and Ch2 observations and simultaneous \emph{HST} and \emph{Spitzer} observations.
The \emph{Spitzer} data allow analysis of continuous temporal variations and constrain longitudinal heterogeneities in the atmospheres, 
while the simultaneous multi-wavelength observations yield pressure-dependent phase shifts and probe atmospheric structures in the vertical direction.

%3. In many visits, light curve shapes are similar but offset between different wavelengths; 
%4. The phase shift combined with the fact that different wavelengths probe different pressures provide information on the vertical longitudinal structure of
%the clouds;

%%5. For objects xxxx, we can measure a phase shift in the fitted curves; For objects xxxx, we can not.

\subsection{Measuring Phase Shifts Between \emph{Spitzer} Ch1 and Ch2 Observations}

\begin{figure*}[ht]
  \begin{center}
    \includegraphics[scale=0.78]{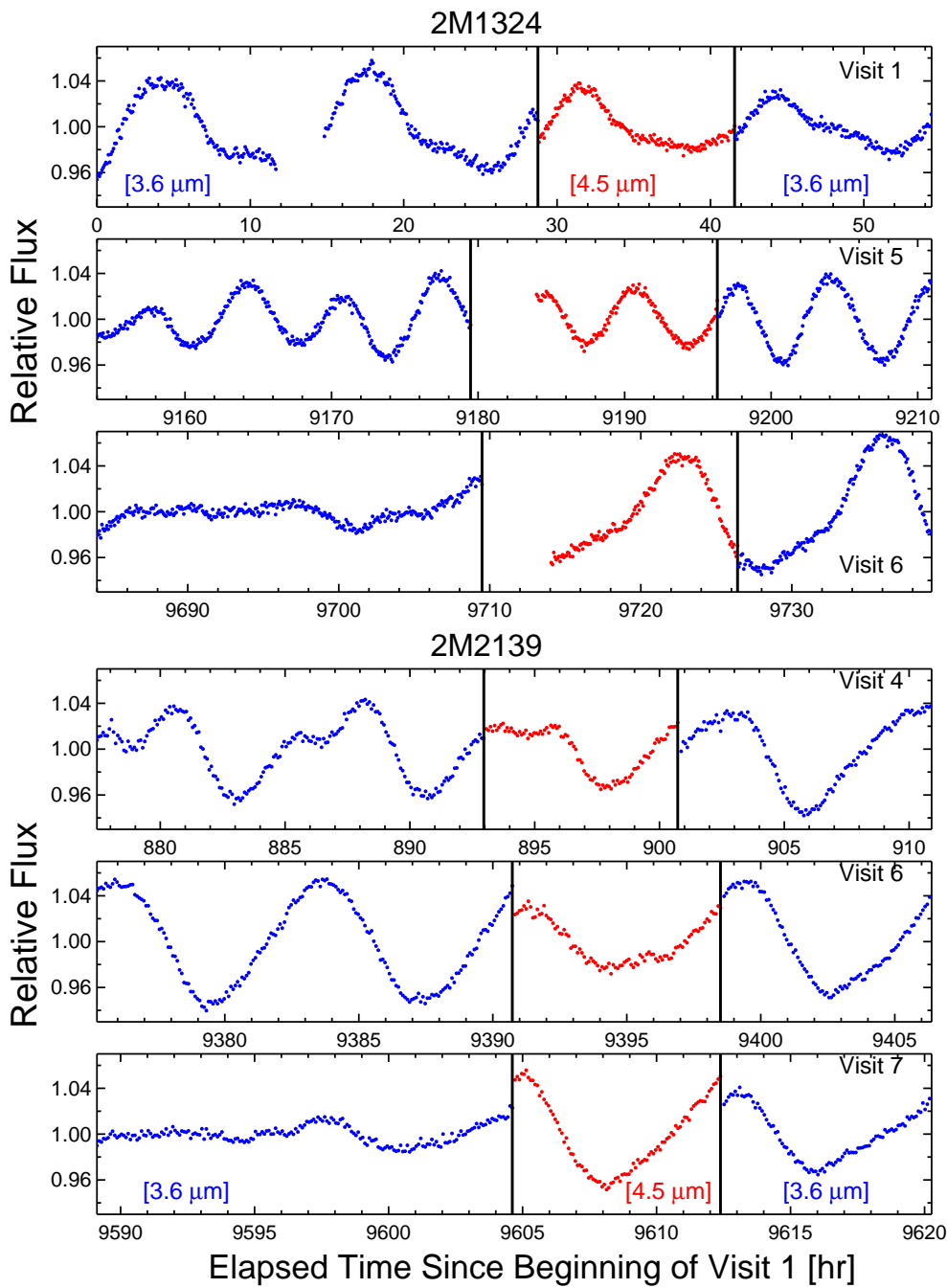}
       \vspace*{5mm}
       \caption{Representative \emph{Spitzer} light curves from three visits of 2M1324 and 2M2139. 
                Each data point is binned from a 5-min chunk of observations. The three parts of light curve in each visit 
                are normalized separately according to their respective median values.}
           \label{fig2replc}
              \end{center}
    \end{figure*}

One of the objectives of our \emph{Spitzer} observations is to measure phase shifts
between Ch1 and Ch2 light curves. In a typical \emph{Spitzer} visit, as shown in Figure~\ref{fig2replc},
we first observe two rotations in Ch1, followed by one rotation in Ch2 and then another rotation in Ch1. 
%Such observation planning allows us to minimize potential additional uncertainties introduced by factors such as 
%light curve evolution and different amplitudes in the two channels.

We employed two methods to determine potential phase shifts between Ch1 and Ch2 observations. The first method derives the 
phase shift from the times at which the target is at maximum or mininum flux levels, $T_{\rm{max}}$ or $T_{\rm{min}}$. 
For visits where we can unambiguously identify the flux maxima and minima in Rotations 2, 3, and 4 (R2, R3, and R4, 
as marked for example in {Figure~\ref{fig1507}}), we first found the midpoint in $T_{\rm{max}}$ or $T_{\rm{min}}$ for R2 and
R4, which are both observed in Ch1. This was achieved by fitting a third-order Fourier function (see Eq. [2]) and then
calculating the maxima and minima from the fits. Next, we calculated the time difference between the $T_{\rm{max}}$ or $T_{\rm{min}}$
in R3 (observed in Ch2) and the midpoint time of R2 and R4. The time difference is expressed as a phase shift
in degrees using the rotation period given in Table~\ref{obs1}. 
Both the time differences in hours and the phase shifts in degrees are given in Table~\ref{shiftmaxmin}.
We excluded the visits where $T_{\rm{max}}$ or $T_{\rm{min}}$ can not be well determined due to light curve evolution
and/or incomplete phase coverage, such as Visit 6 of 2M1324 (see Figure~\ref{fig2replc}). 
%%As the light curves used in the analysis are binned in 5-minute intervals, the time resolution of the measurements is 5 minutes or 0.083 hours. 

%%compare the rotation period determined from R2R4 and auto_correlation

To estimate the uncertainty in the phase shift measurements, we created $1,000$ synthetic light curves by adding random noise to the Fourier fits.
The random noise is drawn from a normal distribution derived from the residuals of the Fourier fit. We performed the same analysis described above on
the synthetic light curves, and the standard deviation of the $1,000$ phase shift measurements were adopted as the uncertainties.

The second method we used to measure phase shifts betweeen light curves taken at different wavelengths 
was to directly cross-correlate the light curves in each visit between the two channels. 
For each visit, we first measured a rotation period by cross-correlating the two rotations observed in Ch1, R2 and R4.  
The shift in time at which the cross-correlation function reaches maximum plus the time difference between the end times of R2 and R4 are equal
to two rotation periods.
Then we cross-correlated R2 (in Ch1) and R3 (in Ch2), and also determined the shift in time corresponding to the peak of the cross-correlation function.  
The phase shift was calculated from the shift in time and the rotation period measured from R2 and R4.
The same analysis was applied to R3 (in Ch2) and R4 (in Ch1). 
We also fit a Gaussian to the cross-correlation function, and the uncertainties 
in the center location of the fitted Gaussian expressed in phase were taken as the uncertainties of the measured phase shifts.
Note that this method does not work well when light curve evolves significantly between consecutive rotations, such as Visit 7 of 2M2139 
(see Figure~\ref{fig2replc}). We list the results from the visits that show little evolution of light curve shape  in Table~\ref{shiftcross}.

\subsection{Measuring Phase Shifts Between \emph{HST} and \emph{Spitzer} Observations}

%For the four objects simultaneously observed with \emph{HST} and \emph{Spitzer}, the \emph{HST} observations 
%have incomplete phase coverages and often overlapped with both Ch1 and Ch2 observations.
%Therefore \emph{HST}-\emph{Spitzer} phase shift measurement requires a different approach 
%than the non-simultaneous but longer and better-sampled Ch1-Ch2 phase shift comparisons.

For the four objects simultaneously observed with \emph{HST} and \emph{Spitzer}, the \emph{HST} observations 
have incomplete phase coverages. %% and often overlapped with \emph{Spitzer} observations in  .
Therefore, \emph{HST}-\emph{Spitzer} phase shift measurement requires a different approach 
than the non-simultaneous but longer and better-sampled Ch1-Ch2 phase shift comparisons.

\begin{figure*}[ht]
  \begin{center}
    \includegraphics[scale=0.76, angle=90]{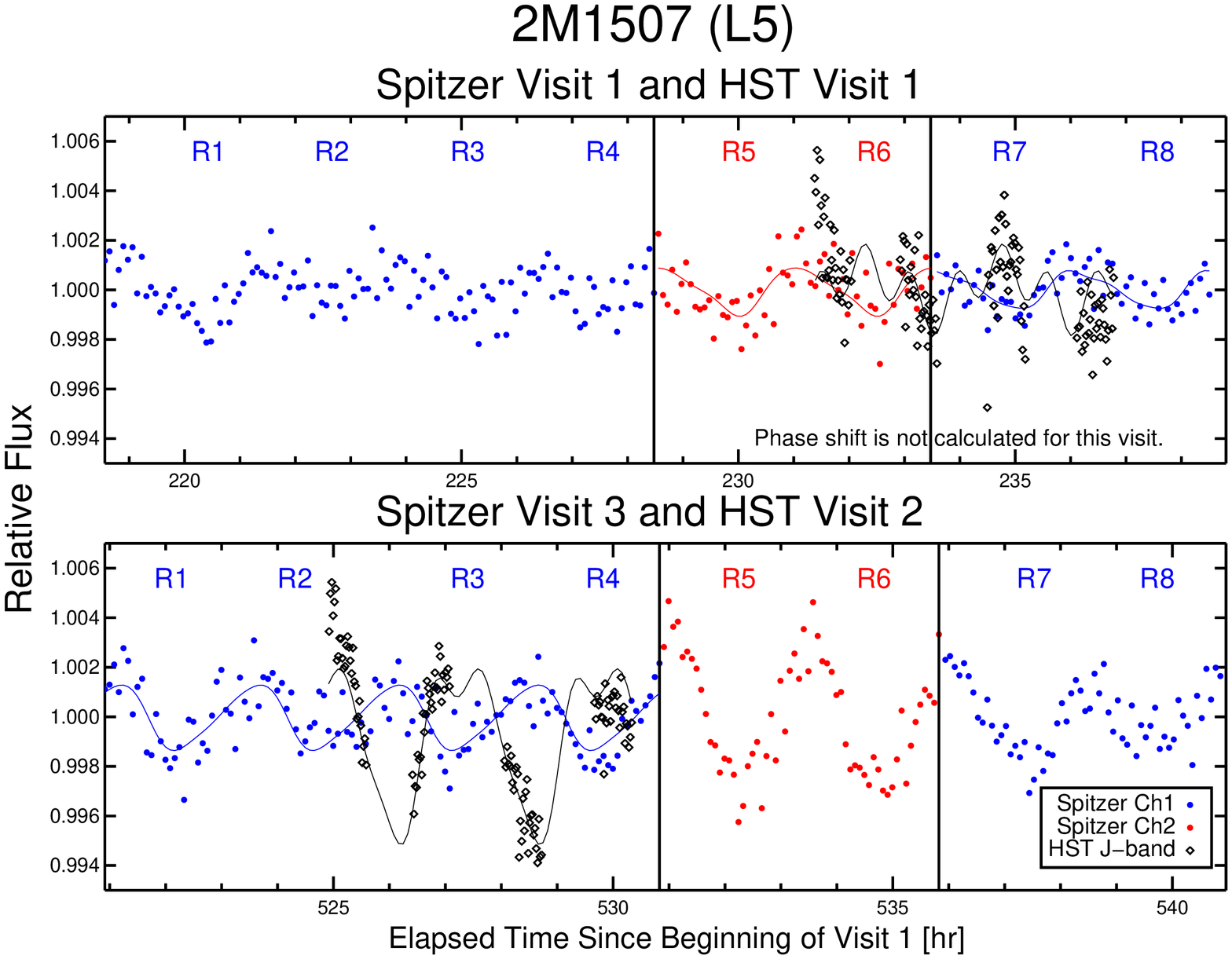}
       \vspace*{2mm}
       \caption{Simultaneous \emph{HST} $J$-band and \emph{Spitzer} light curves of 2M1507. 
               The black solid curves are Fourier fits to the \emph{HST} $J$-band light curves. 
               The blue and red solid curves are Fourier fits to the \emph{Spitzer} Ch1 and Ch2 light curves, respectively. 
               The amplitude of the $J$-band light curves is increased by a factor of 1.5 for display purpose.
                }
           \label{fig1507}
              \end{center}
    \end{figure*}

\begin{figure*}[ht]
  \begin{center}
    \includegraphics[scale=0.76, angle=90]{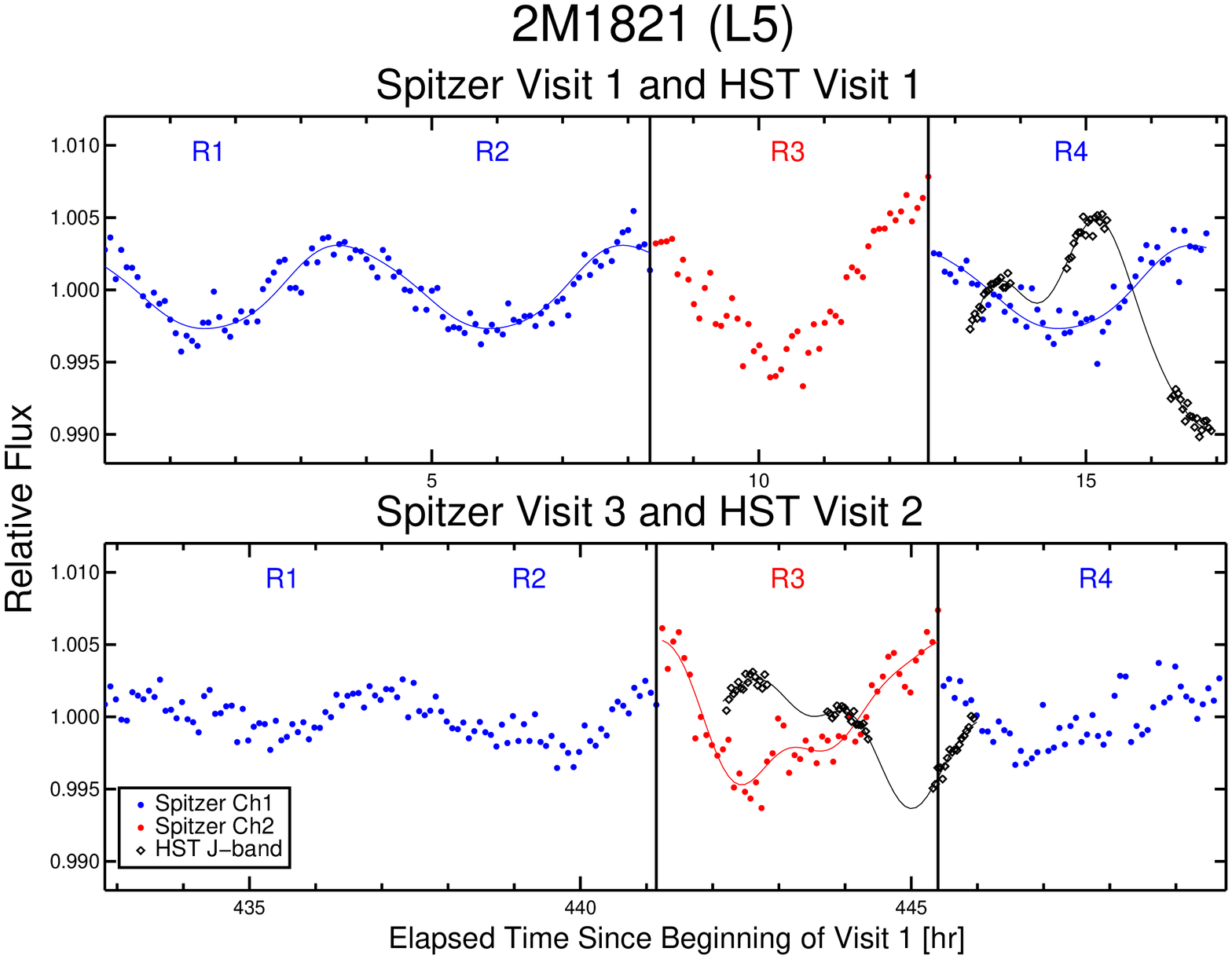}
       \vspace*{2mm}
       \caption{Simultaneous \emph{HST} $J$-band and \emph{Spitzer} light curves of 2M1821. 
               The black solid curves are Fourier fits to the \emph{HST} $J$-band light curves. 
               The blue and red solid curves are Fourier fits to the \emph{Spitzer} Ch1 and Ch2 light curves, respectively. 
               The amplitude of the $J$-band light curve is decreased by a factor of 1.5 for display purpose.
                }
           \label{fig1821}
              \end{center}
    \end{figure*}

\begin{figure*}[ht]
  \begin{center}
    \includegraphics[scale=0.76, angle=90]{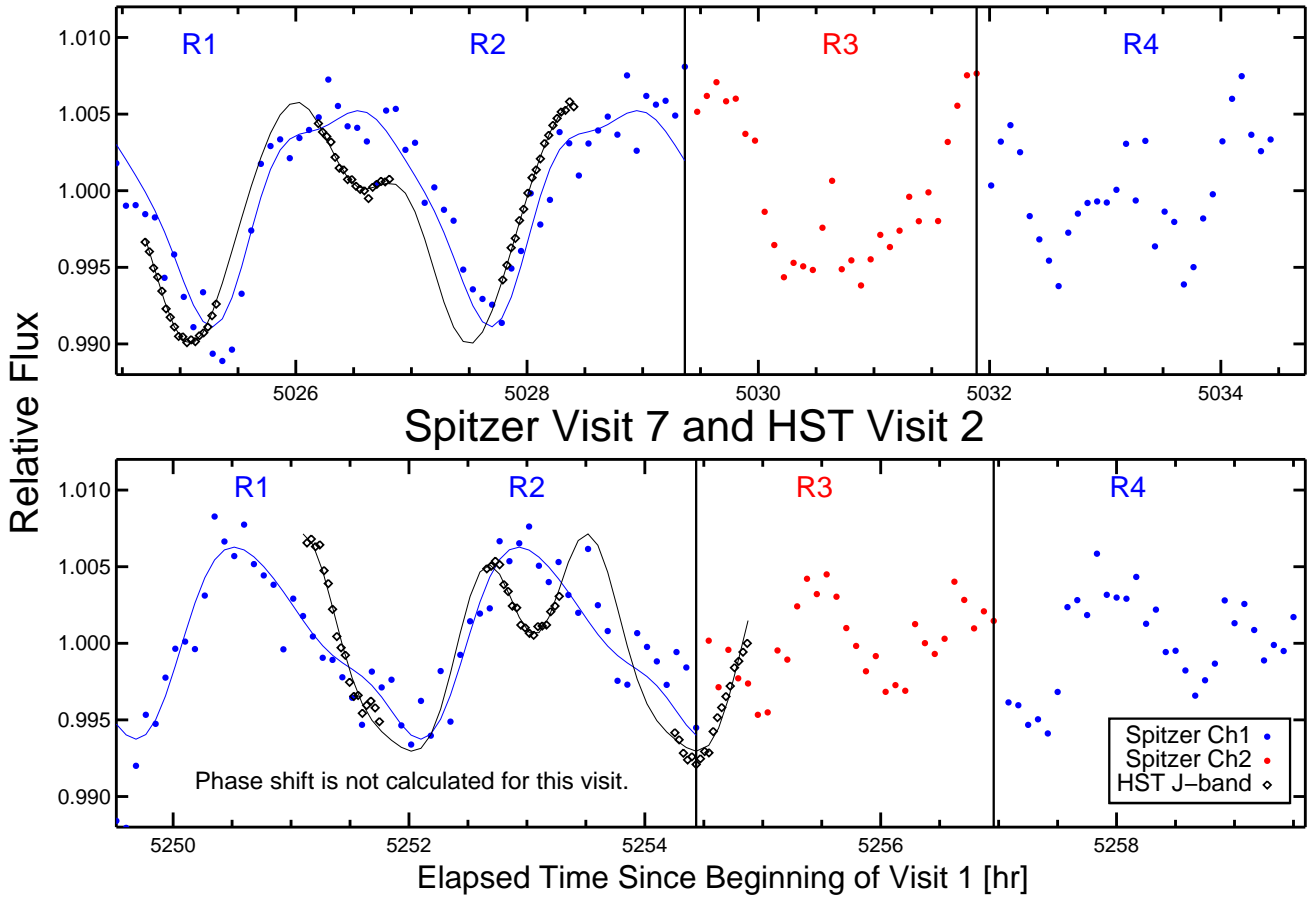}
       \vspace*{2mm}
       \caption{Simultaneous \emph{HST} $J$-band and \emph{Spitzer} light curves of SIMP0136. 
               The black and blue solid curves are Fourier fits to the \emph{HST} $J$-band and \emph{Spitzer} Ch1 light curves, respectively. 
               The amplitude of the $J$-band light curves is decreased by a factor of 3.5 for display purpose.
               }
           \label{fig0136}
              \end{center}
    \end{figure*}

\begin{figure*}[ht]
  \begin{center}
    \includegraphics[scale=0.76, angle=90]{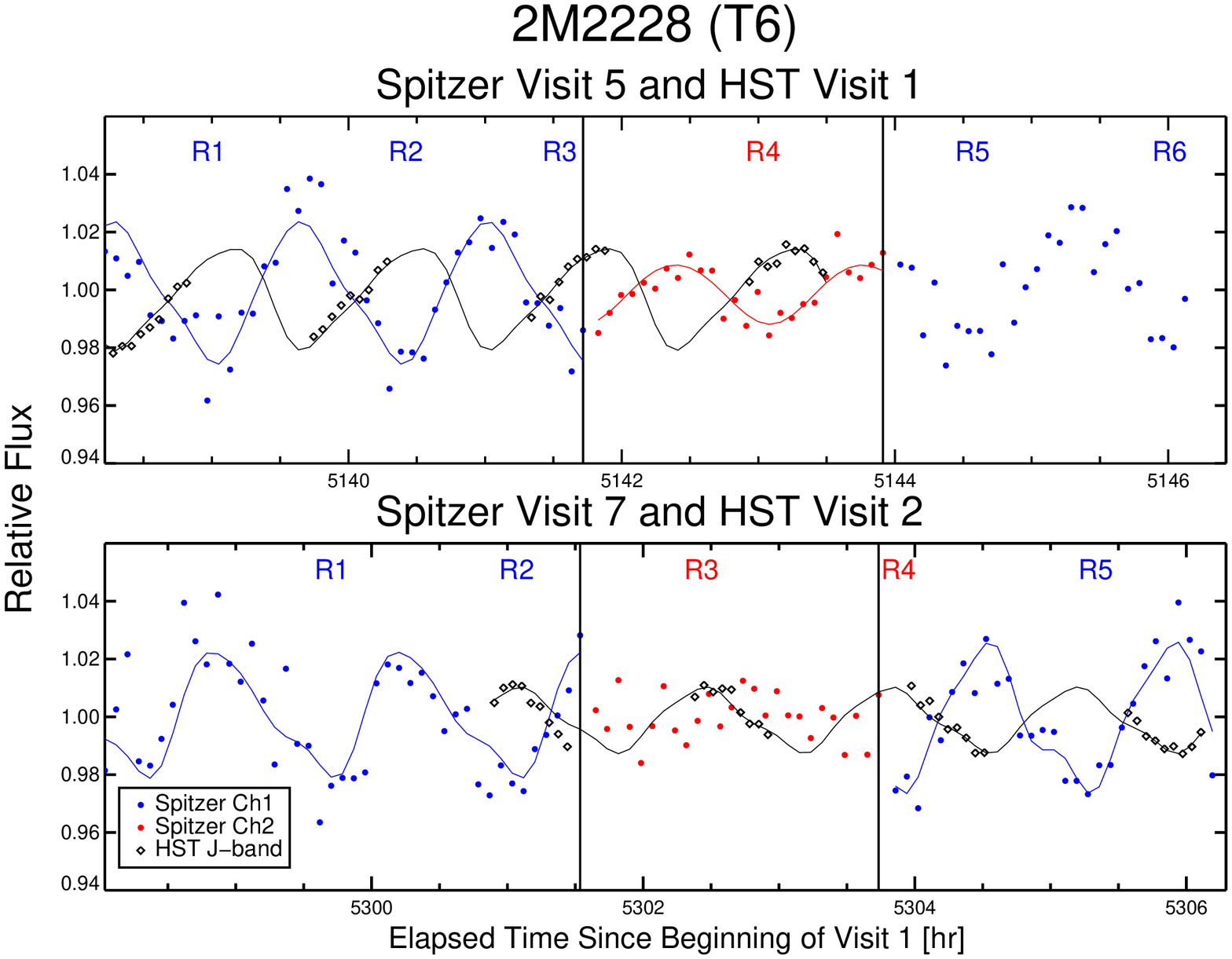}
       \vspace*{2mm}
       \caption{Simultaneous \emph{HST} $J$-band and \emph{Spitzer} light curves of 2M2228. 
               The black solid curves are Fourier fits to the \emph{HST} $J$-band light curves. 
               The blue and red solid curves are Fourier fits to the \emph{Spitzer} Ch1 and Ch2 light curves, respectively. 
               The amplitude of the $J$-band light curve is increased by a factor of 1.5 for display purpose.
                }
           \label{fig2228}
              \end{center}
    \end{figure*}

We compare \emph{HST} $J$-band light curves with \emph{Spitzer} light curves of the four objects in {Figures~\ref{fig1507}--\ref{fig2228}}. 
{The $J$-band light curves were} calculated by integrating the \emph{HST}/WFC3 grism spectra over the 2MASS $J$-band relative spectral 
response curve \citep{cohen2003}. To be consistent with the timestamps of the \emph{Spitzer} observations, 
all \emph{HST} timestamps were converted from Modified Julian Dates to barycentric Modified Julian Dates,
using the IDL routine $barycen.pro$. The mid-time of each \emph{HST} exposure is used as the time of observation. The light curves
from the two instruments are normalized by their respective median values.

To quantify the phase shifts, we fit separate third-order Fourier functions (see Eq.[2]) to the simultaneous \emph{HST} J-band and 
\emph{Spitzer} Ch1 or Ch2 light curves. The rotation periods in the Fourier functions were fixed to the values listed in Table~\ref{obs1}.
We used the best-fit parameters of the Fourier fits to calculate the phases of the first- and second-order Fourier components and
then, from their difference, the corresponding phase shifts. 

Given the high precision of the \emph{HST} data, the main source of uncertainty in the measured
phase shifts is from the Fourier fits to the \emph{Spitzer} light curves. We estimated the amplitude of noise by fitting a Gaussian to 
the residuals of the Fourier fit, and created a synthetic light curve by adding random noise drawn from the fitted Gaussian distribution to the Fourier fit. 
We then measured the phase shift in the synthetic light curve in the same fashion as the observed light curve. 
This procedure was repeated 1,000 times for each segment of \emph{Spitzer} light curve
analyzed and the standard deviation of the 1,000 measurements was taken as the uncertainties of the phase shift.
We report the phase shifts and associated uncertainties for first- and second-order Fourier components in Table~\ref{shift1}.
%%Due to light curve evolution, The measurements from \emph{Spitzer} Visit 7 of SIMP0136 
%%The Fourier fit to the \emph{HST} light curve in Visit 1 of 2M1507 is poor.
In the following discussion, we excluded measurements from two visits, \emph{Spitzer} Visit 7 of SIMP0136 and Visit 1 of 2M1507, 
because for Visit 7 of SIMP0136 (bottom panel of Figure~\ref{fig0136}), light curve evolution makes the measurement ambiguous, 
and for Visit 1 of 2M1507 (top panel of Figure~\ref{fig1507}), the Fourier fit to the \emph{HST} data is poor.

\section{RESULTS}

\subsection{Variability in all sources and all spectral bands}

We found that all six sources in our sample are variable in both IRAC channels in all \emph{Spitzer} visits. 
Most light curves are not strictly sinusoidal and can evolve on timescales as short as one rotation period, 
as shown in {Figures~\ref{fig2replc}--\ref{fig2228}}. 
The simultaneous \emph{HST} observations of four short-period sources also exhibit flux variations in the integrated light curves of 
different narrow spectral bandpasses (Figures~\ref{narrow1507}--\ref{narrow2228}).
%%More detailed discussion on the evolution of light curve shape and amplitude will be presented in a separate paper (Apai et al. 2016, in preparation).

\subsection{Rotation Periods}

For most of our targets our Spitzer data represents the longest continuous, high-precision light curves. As such, these data can place powerful constraints 
on the rotational period of the objects. In order to examine periodicity and estimate the rotational period, we calculated the auto-correlation functions 
for each visit for the Ch1 observations (approximately three rotations of the targets per visit sampling a time interval corresponding to approximately 
four rotation periods, see Figure~\ref{fig1schedule}). We used the standard $\mathrm{IDL}$ script $a_correlate.pro$ for the auto-correlation analysis 
and identified the first peak (corresponding to non-zero shifts) in the autocorrelation function for each visit. 

For three of our sources (SIMP0136, 2M2139, and 2M2228), the autocorrelation function was well-defined in most visits and the peaks identified were consistent. 
{Examples are shown in Figure~\ref{autocorr}.}
For these objects we adopted, as rotation periods, the mean of the peak auto-correlation values and, as the uncertainties of the rotation periods, 
the standard deviation of the peak locations in the auto-correlation functions.

For the other three targets, however, the light curve evolution was so significant that no peak in the auto-correlation function emerged consistently 
among the eight visits. For these three objects, therefore, the auto-correlation analysis did not yield reliable rotation period estimates. 
For these source we adopted previous rotation estimates by \citet{metchev2015}, primarily based on Fourier analyses, which are qualitatively consistent 
with our light curves. The measured and adopted rotation periods are listed in Table~\ref{obs1}.

%We performed auto-correlation analysis on the \emph{Spitzer} light curves [more details here]. We accurately measured rotation periods of
%three objects, SIMP0136, 2M2139, and 2M2228. 
%Previously, \citet{buenzli2012} measured a rotation period of $1.38 \pm 0.03$ hr for 2M2228,
%and \citet{apai2013} estimated the rotation periods for SIMP0136 and 2M2139 to be $2.39 \pm 0.07$ hr and $7.83 \pm 0.1$ hr, respectively.
%With much longer observing time and better phase coverage than previous monitoring of the three sources, 
%our measurements are consistent with previous estimates and are more reliable. 
%For the other three objects, the light curve evolution limited the analysis, so we adopted the results from \citet{metchev2015}.
%Fourier fits with rotation periods fixed at the adopted values from literature are generally consistent with the \emph{Storms} data.

\subsection{Phase Shift Between \emph{Spitzer} Ch1 and Ch2 Observations}

We utilized two different methods to measure phase shifts between light curves observed in the two IRAC channels. While both methods are somewhat limited by
light curve evolution on short timescales, our measurements showed that there is no detectable phase shift between the Ch1 and Ch2 light curves. 
As listed in Tables~\ref{shiftmaxmin} and \ref{shiftcross}, the majority of measurements are within 1- or 2-$\sigma$ limit of 0$^\circ$, with a typical
upper limit of $15^\circ$.

\subsection{Phase Shift Between \emph{HST} and \emph{Spitzer} Observations}

Simultaneous \emph{HST} and \emph{Spitzer} observations of the four targets from the \emph{Storms} program provide the most comprehensive 
brown dwarf monitoring dataset, with each object covered for at least six \emph{HST} orbits. Previously, only \citet{buenzli2012} have studied 2M2228 with
simultaneous \emph{HST} and \emph{Spitzer} observations, which overlapped for two \emph{HST} orbits. 
With our unique dataset, we found that SIMP0136 (Figure~\ref{fig0136})
shows a phase shift of $\sim 30^\circ$ between \emph{HST} $J$-band and \emph{Spitzer} light curves, 
while the other three objects (2M1507, 2M1821, and 2M2228) all show substantial phase shifts within $10-20^\circ$ of $180^\circ$ 
(Table~\ref{shift1}). 

\subsection{Phase Shift Between \emph{HST} Narrow-Band Light Curves}

\citet{buenzli2012} discovered phase shifts among several integrated light curves of characteristic wavelength regions 
in the \emph{HST} and \emph{Spitzer} observations of 2M2228 obtained in July, 2011. The selected bandpasses, including the $J$- and $H$-band windows 
as well as water and methane absorption, probe different pressure levels in the atmosphere, and \citet{buenzli2012} found that the phase shifts 
increase monotonically with decreasing pressure level, indicating vertical atmospheric structures. 

We applied the same analysis procedure as described in \citet{buenzli2012} to the \emph{Storms} observations of 2M2228, and found that
the light curves of various narrow bandpasses still display phase shifts even after two years, corresponding to over 12,000 rotations. 
We fit sine waves to the narrow-band light curves and compared
the phases of the sine waves. As shown in Figure~\ref{narrow2228}, with respect to the 
$J$- and $H$-band light curves, the water and methane bands shows phase shifts close to 180$^\circ$ in both \emph{HST} visits. 
We extended this analysis to all four \emph{Storms} targets observed with \emph{HST}. 
The T2 dwarf SIMP0136 (Figure~\ref{narrow0136}) shows little phase shift, typically $4^\circ\pm 2^\circ$. We discuss this finding in the next section.

\section{DISCUSSION}

\subsection{Atmospheric Models and Pressure-dependent Flux Contribution}

The greatest advantage of multi-band monitoring of brown dwarfs is that emergent fluxes at different wavelengths probe different depths in the atmospheres.
To investigate the characteristic pressure levels that different wavelength regions probe, we employ state-of-the-art radiative transfer and 
atmospheric chemistry models and calculate relative flux contributions from each model pressure levels.

We first performed least-squares fits to available \emph{HST} grism spectra to find the best-fit atmospheric model for each target. 
The atmosphere models we use are the most up-to-date versions of those published in \citet{saumon2008}.
Besides the four targets observed with the \emph{Storms} program,  we also used \emph{HST}/WFC3 G141 spectra of 2M2139 from \citet{apai2013}. 
The observations were resampled 
according to the model spectral grid. The model spectra were normalized to match with the flux peak in the $J$-band between 1.25 and 1.28 $\mu$m.
The observed and the best-fit model spectra are shown in Figure~\ref{modelfits}, and the best-fit model parameters are 
given in Table~\ref{modelparam}. 
The observed spectrum beyond 1.55 $\mu$m for 2M1821 is contaminated by a background object and thus not well fit by the model.
As 2M1324 has not been observed with the WFC3 G141 grism, we use the best-fit model for the other T2 dwarfs.

To compute the contribution functions, we first converged a standard radiative-convective equilibrium atmosphere thermal structure model following the approach of \citet{saumon2008}. 
Since the {time} of that model description, there have been substantial updates to the opacity database employed, which {were utilized in the present analysis and}
will be described more fully in an upcoming paper (Marley et al., in prep.). 
Once the model converged, a temperature perturbation was iteratively applied to quarter-scaleheight subregions of the atmosphere. 
Given this new, artificial temperature profile a new emergent spectra was computed. The perturbation was then removed, a new perturbation applied to the next overlying region, 
and the process repeated. By {computing the ratio of} each perturbed thermal emission spectrum to the baseline case, 
the sensitivity of each spectral region to temperature perturbations at depth could be computed. 

%%Note, however, that since the cloud opacity was not altered this approach strictly applies only to changes in atmospheric temperature alone. 
%%However, this approach does make clear the atmospheric region which the spectra at a given wavelength are most sensitive to. Changes to the cloud opacity at or above 
%%this level are likely to substantially alter the emergent flux whereas changes to the underlying cloud opacity are less likely to be as significant.

We repeated this procedure for 37 model pressure levels that cover pressures from $1.8 \times 10^{-4}$ bars to $\sim$ 23 bars, and essentially obtained 
the relative flux contributions from a range of pressures at the wavelengths covered by our \emph{HST} and \emph{Spitzer} observations
(left panels of Figures~\ref{contrifunc1700}--\ref{contrifunc950}). 

%%For a given atmosphere model, to determine the relative flux contributions from a certain pressure level at the wavelengths we observed, 
%%we first introduced a 50 K thermal perturbation in the temperature profile at the specific model pressure grid. Each model pressure grid corresponds to 
%%one quarter pressure scale height. Then we generated a new spectrum from the perturbed model. 
%By taking the ratio of the perturbed model spectrum over the original model spectrum at each wavelength, we obtained the relative flux variations due to the temperature perturbation. 
%%Note that the relative flux contribution at different pressures should be compared only for the same wavelength. Comparison of the percentages between
%%two different wavelengths is not meaningful.

To find the characteristic pressure level where most of the flux emerges from for a spectral bandpass of interest, 
we integrated the relative flux contributions over the wavelength bandpass and calculated a cumulative flux contribution function starting from 
the top of the atmosphere. We identified the two pressure levels between which the cumulative flux reaches 80\% of the total flux and determined the exact
pressure level by linear interpolating between the two model pressure levels. The 80\%-cumulative-flux pressure levels were calculated for 8 spectral 
bandpasses, including 5 \emph{HST} narrow bands as well as 2MASS $J$-band and \emph{Spitzer} Ch1 and Ch2. The results are listed in Table~\ref{modelparam}
and also shown in the right panels of Figures~\ref{contrifunc1700}--\ref{contrifunc950}.

{We stress that the approach adopted here, although based on state-of-the-art atmosphere models, offers only a limited tool for identifying the specific pressure levels where modulations are introduced:
In our approach of modifying the temperature but not the cloud opacity, the results strictly apply only to changes in atmospheric temperature {\em alone}. However, our model does make clear the atmospheric region which the spectra at a given wavelength are most sensitive to.
Changes to the cloud opacity at or above the pressure levels we identify for each observations are likely to substantially alter the emergent flux whereas changes to the underlying cloud opacity (below the 80\% contribution levels) are less likely to be as significant.} 

{A good example of how high-altitude clouds or haze layer can dramatically modulate the pressure levels probed in the infrared is provided by \emph{Cassini} observations of hot spots in Jupiter’s equatorial regions. \citet{choi2013} presents multi-band \emph{Cassini} imaging monitoring of the hot spots and plumes that are seen to co-evolve near the jovian equator. The hot spots are interpreted as cloud clearing in the ammonia cloud deck, which may be rapidly obscured or revealed by high-altitude clouds. In this case, for example, the infrared observations probe deep in the atmosphere when cloud opacity is {\em absent}, but are limited to the top of the atmosphere when opacity is introduced by high-level clouds.}

\subsection{Heterogeneous Upper-atmosphere in All Objects}

All six sources in our sample exhibit flux variations in both \emph{Spitzer} channels during all 8 visits, regardless of spectral type and rotation period.
We found that the derived characteristic model pressures above which 80\% of the fluxes emerge from at Ch1 and Ch2 are less than 2 bars for the two L5 
dwarfs and less than 1 bar for the T dwarfs, indicating heterogeneity in the upper atmospheres of all our targets.

Interestingly, the relative flux contributions (Figures~\ref{contrifunc1700}--\ref{contrifunc950}) show that in the model atmosphere 
of a T2 dwarf most of \emph{Spitzer} Ch1 flux is from lower pressure levels than the Ch2 flux, while the case is reversed in the L5 and 
T6 models. 

For 2M2228 (T6), shown in top left panel of Figure~\ref{contrifunc950}, flux between 1.1 and 1.7 $\mu$m emerges from a wide range of pressures levels.
For example, the narrow $J$-band flux is mostly from $\sim$ 7.5 bars, while the flux of the $1.35-1.43~\mu$m water band primarily comes from $\sim$ 2 bar.
In contrast, the L5 and T2 dwarfs emit most of the flux between 1.1 and 1.7 $\mu$m from a smaller range of pressures, which are around 6.6-4.3 bars 
for the L5 targets and 8.1-4.1 bars for the T2 targets.

\subsection{Observed Phase Shifts}

Between \emph{Spitzer} Ch1 and Ch2 light curves, we have found no detectable phase shifts from analyses using two different methods. 
According to the relative flux contributions derived from the models (Figures~\ref{contrifunc1700}--\ref{contrifunc950}), most of the observed flux 
in both \emph{Spitzer} channels comes from a relatively narrow range of pressure levels in the upper atmospheres, e.g., between 1 and 2 bars in Ch1 
for the L5 model (Figure~\ref{contrifunc1700}), and between 0 and 0.3 bars in Ch2 for the T2 model (Figure~\ref{contrifunc1400}).

For all four targets that have simultaneous \emph{HST} and \emph{Spitzer} observations, we have detected phase shifts between $J$-band and Ch1/Ch2
light curves (Table~\ref{shiftcross}). 
The rotation rates of our objects range from 1.37 hr (2M2228) to 4.2 hr (2M1821).
With the small sample size, the phase shifts display no obvious correlation with rotation periods.
For example, SIMP0136 (T2) and 2M1507 (L5) have similar rotation periods, but their phase shifts are very different.
SIMP0136 (T2) exhibits $\sim 30^\circ$ phase shift, while the phase shifts of the mid-L and late-T dwarfs are all centered around 180$^\circ$. 
This might hint that L/T transition objects have peculiarities in their atmospheric properties with respect to regular L and T dwarfs,
resulting in different phase shifts between the \emph{HST} and \emph{Spitzer} light curves.

\subsection{Correlation Between Pressure Levels and Phase Shifts}

Simultaneous observations in multiple bandpasses probes different pressure levels in the atmosphere, providing vital information on the
atmospheric properties in the vertical direction.
With the \emph{Storms} data set, we measured phase shifts between light curves of four narrow bands between 1.1 and 1.7 $\mu$m 
and two \emph{Spitzer} broad bands centered at 3.6 and 4.5 $\mu$m.
Here we explore the correlation between phase shifts of the six bandpasses and
the characteristic model pressure levels from which most flux originates for those bandpasses. 

\citet{buenzli2012} first detected pressure-dependent phase shifts in 2M2228, with lower pressure levels showing larger phase shifts. 
In Figure~\ref{shftpressure2228}, we examine again the phase shifts as a function of model pressures for 2M2228 observed two years later.
We find that the phase shifts measured among \emph{HST} narrow bandpasses in this work are consistent with those reported by \citet{buenzli2012}.
Our measured phase shifts between the $J$- and $H$-band light curves are $-5^\circ \pm 2^\circ$ and $-8^\circ \pm 2^\circ$ for two visits, respectively 
(Figure~\ref{narrow2228}), while \citet{buenzli2012} found it to be $15^\circ \pm 2^\circ$.
The sine wave fits to the $1.35-1.43~\mu$m (water) and $1.62-1.69~\mu$m (water and methane) bandpasses are not as good as $J$- and $H$-bands, and 
the phase shifts of those two bandpasses with respect to the narrow $J$-band are generally close to 180$^\circ$, consistent within errors 
with the results of \citet{buenzli2012}. {(Note that a 180$^\circ$ phase shift could be equivalent to an anti-correlation of fluxes at different
wavelengths with no phase shift.)}
The main difference between the results of \citet{buenzli2012} and this work lies in the phase shift between the narrow $J$-band and the \emph{Spitzer} 
channels, which \citet{buenzli2012} measured to be $118^\circ \pm 7^\circ$, compared to our measurement of around 160$^\circ$.

Our Fourier-based phase shift measurements between different wavelength bands are separated into two distinctive groups 
in terms of the pressure levels probed by the bandpasses.
For each target (left panels of Figures~\ref{shftpressure2228}--\ref{shftpressure1821}), the light curves probing deeper ($\gtrsim$ 4 bars) in the atmosphere 
are generally in phase, while the light curves probing the upper atmospheres ($\lesssim$ 4 bars) display similar phase offsets. 
For the L and the L/T dwarfs, the two groups of light curves appear to probe pressure levels separated approximately by 
the radiative-convective boundary calculated from the model atmospheres. For the T6 object, the radiative-convective boundary is deeper in the atmosphere than
where our available wavelength bands can probe.

Compared with the locations of the condensate clouds in the models (right panels of Figures~\ref{shftpressure2228}--\ref{shftpressure1821}), 
the higher pressure region ($\gtrsim$ 4 bars) probed by our observations generally coincides with the dense part of the cloud layers.
Variable cloud thickness could explain the observed flux variations in the light curves that probe  higher pressure levels in the atmosphere. 
On the other hand, the pressure levels that the \emph{Spitzer} channels probe have very low condensate density.
Along with the phase differences, this might indicate that different sources of modulation are in play in the two regions. 
The transition between the two groups of light curves is very abrupt in pressure, indicating a break or discontinuity between
the two pressure regions.
A separate patchy cloud deck at the lower pressure levels could cause the flux variations probed by the \emph{Spitzer} bands.

\section{SUMMARY}

We monitored the light curves of two mid-L dwarfs, three L/T transition dwarfs, and one late-T dwarf over the course of 20 months. 
We cover at least 24 rotations in Spitzer observations for all six targets and sample at least six rotations with time-resolved \emph{HST}
spectroscopy for four objects. The key results of our study are:
   \begin{itemize}
    \item  All six targets are variable and exhibit light curve evolution over timescales of their rotation periods, ranging from 1.4 h to 13 h.
      For each object, we find variations in every wavelength band observed, demonstrating that the photospheres of all six objects are heterogeneous.

     \item  For three objects, we accurately determine rotation periods. For the other three, we adopt estimates of rotation periods from \citet{metchev2015}
and find those values consistent with much longer observations in this work.

     \item We use state-of-the-art radiative transfer and atmospheric chemistry models to determine the flux contribution of each pressure layer 
           to the spectral bands studied.
           Our observations probe model pressure levels between $\sim$8.1 to $\sim$0.2 bars, using light curves obtained in 7 different 
           bandpasses for four objects.  
%%For the two long-period objects, we have \emph{Spitzer coverage probing pressures from $\sim$1.0 to $\sim$0.29 bar.

     \item  We use two different methods to assess the phase shifts between the spectral bands studied for our objects.
No phase shift is found in any of six objects between \emph{Spitzer} Ch1 and Ch2 light curves. Both channels probes a narrow range of layers 
high in the atmospheres ($\lesssim$ 3 bars for the L dwarfs and $\lesssim$ 1 bar for the T dwarfs). %%as shown by the contribution functions.

     \item  We detect phase shifts between \emph{HST} $J$-band and \emph{Spitzer} light curves for all four objects simultaneously observed 
by the two observatories. From the limited sample, the phase shift between the \emph{HST} and \emph{Spitzer} data does not show correlation with rotation period.

     \item  The \emph{HST} $J$-band and Spitzer light curves of SIMP0136 (T2) shows a small phase shift ($\sim 30^{\circ}$), 
while for the other three targets such phase shifts are close to $180^{\circ}$. This might indicate 
that the L/T transition objects have peculiar atmospheric properties, compared to regular L and T dwarfs. 
More simultaneous multi-band observations are needed to check whether such different behavior in phase shifts are common for L-T transition dwarfs.

     \item  For 2M2228 (T6), phase shifts among narrow \emph{HST} bands and \emph{Spitzer} Ch2 persist between two sets of observations 
separated by two years, equivalent of thousands of rotations. The phase shifts among \emph{HST} narrow bandpasses are generally consistent 
with that reported by \citet{buenzli2012}, while the phase shift between the narrow $J$-band and \emph{Spitzer} Ch2 is different from that measured in 2012.
No \emph{HST} narrow-band phase shifts are found for 2M1507 (L5), 2M1821 (L5), and SIMP0136 (T2).

%%Most of the \emph{HST} narrow-band fluxes for the T6 dwarf emerge from a larger range of pressure levels in the atmosphere as compared to the L5 and T2 dwarfs,
%%and thus these narrow-bands probe 
%%     For the wavelength coverage probed by the \emph{HST}/WFC3 G141 grims, the relative flux contribution functions 
%%%%show that for the L5 and T2 dwarfs most of the flux emerges from a smaller range of pressure levels in the atmosphere as compared to a T6 dwarf.

    \item For the four sources with \emph{HST} and \emph{Spitzer} light curves, we identify a clear difference between higher pressures 
($\gtrsim$ 4 bars) and lower pressures ($\lesssim$ 4 bars), 
visible in differences in the light curve shape and Fourier-based phase shifts. 
        The pressure range that separates the two groups of light curves appears to be close to the estimate radiative/convective boundary 
for the mid-L and the L/T dwarfs, but for the T6 dwarf it appears to be significantly lower than the radiative/convective boundary.

  \item We attribute the modulations introduced in the deeper atmosphere to cloud thickness variations occurring at or near the densest parts 
of the condensate clouds probes.  %% (Mg2SiO4). 
     In contrast, the \emph{Spitzer} bands probe pressures where the \citet{saumon2008} cloud models predict only low condensate volume 
mixing ratios, which are unlikely to account for the variations observed. The two groups of light curves and the abrupt transition 
between the phases of the two pressure regions may indicate that another heterogeneous cloud layer at much lower pressures could be responsible.  
Our results show evidence for a possible two-component vertical cloud structure, but not for more components, providing new insights and constraints 
for vertical cloud models.

%% \item Our observations therefore show the presence of two distinct heterogeneous cloud layers in all targeted brown dwarfs: one at or near 
%%the silicate condensate cloud base and one at much lower pressures, at or above the approximate cloud top.

   \end{itemize}

%High-level conclusion here.

%% Our observations provide valuable test cases for upcoming multi-dimensional modeling of brown dwarfs.

%%     \item  Ch1 and Ch2 amplitude qualitative comparison.
\acknowledgments 

This work is part of the Spitzer Cycle-9 Exploration Program, Extrasolar Storms. This work is based in part 
on observations made with the Spitzer Space Telescope, which is operated by the Jet Propulsion Laboratory, 
California Institute of Technology under a contract with NASA. Support for this work was provided by NASA 
through an award issued by JPL/Caltech. 

Support for \emph{HST} GO programs 13176  was provided %by NASA through a grant from the Space Telescope Science Institute, which is operated by the Association 
of Universities for Research in Astronomy, Inc., under NASA contract NAS5-26555. We acknowledge the outstanding 
help of Patricia Royle (STScI) and the Spitzer Science Center staff, especially Nancy Silbermann, for 
coordinating the HST and Spitzer observations. 

This research has benefitted from the SpeX Prism Spectral Libraries, maintained by Adam Burgasser at \url{http://pono.ucsd.edu/~adam/browndwarfs/spexprism} and
the M, L, T, and Y dwarf compendium housed at DwarfArchives.org.

%E.B. is supported by the Swiss National Science Foundation (SNSF). 

\bibliographystyle{apj}
\bibliography{hao}

\pagestyle{empty}

%%%%%%%%%%%%%%%%%%%%%%%%%
%TABLES
%%%%%%%%%%%%%%%%%%%%%%%%%
%\input{table1.tex}

%\begin{landscape}
\begin{center}
\begin{table*}
  \caption{Targets}
   \label{obs1}
%\rotate
%\centering
\resizebox{\textwidth}{!}{%
%\scriptsize
%  \begin{sideways}
  \begin{tabular}{ cccccccc}

    \hline
     Target            &     SIMP0136        &        2M1324       &        2M1507       &        2M1821       &       2M2139        &      2M2228         \\
    \hline
    \hline
    2MASS \#           & J01365662+0933473   &   J13243553+6358281 &  J15074769-1627386  & J18212815+1414010   & J21392676+0220226   &  J22282889-4310262  \\
    R.A.(J2000)        &  01 36 56.62        &     13 24 35.538    &    15 07 47.693     &    18 21 28.153     &   21 39 26.769      &    22 28 28.894     \\
    Dec.(J2000)        &  +09 33 47.3        &     +63 58 28.15    &    -16 27 38.62     &    +14 14 01.04     &  +02 20 22.70       &   -43 10 26.27       \\
    SpType             &       T2            &         T2          &          L5         &           L5        &       T2            &        T6           \\ 
    J [mag]            &    13.46 $\pm$ 0.03     &   15.596 $\pm$ 0.067    &   12.830 $\pm$ 0.027    &   13.431 $\pm$ 0.024    &   14.710 $\pm$ 0.003    &  15.662 $\pm$ 0.073     \\ 
    H [mag]            &    12.77 $\pm$ 0.03     &   14.576 $\pm$ 0.056    &   11.895 $\pm$ 0.024    &   12.396 $\pm$ 0.019    &   14.16 $\pm$ 0.05      &  15.363 $\pm$ 0.117     \\ 
    K [mag]            &    12.562 $\pm$ 0.024   &   14.06 $\pm$ 0.06      &   11.312 $\pm$ 0.026    &   11.650 $\pm$ 0.021    &   13.58 $\pm$ 0.04      &  15.296 $\pm$ 0.206     \\   % JHK from SIMBAD
  $[3.6]$ [mag]        &    11.359           &   12.896            &   10.370            &   10.608            &   12.389            &  14.519             \\ 
  $[4.5]$ [mag]        &    10.948           &   12.312            &   10.394            &   10.469            &   11.716            &  13.281             \\ 
%%  [3.6] (mag)        &          (    )     &          (     )    &          (     )    &          (     )    &          (     )    &        (    )       \\ 
%%  [4.5] (mag)        &          (    )     &          (     )    &          (     )    &          (     )    &          (     )    &        (    )       \\ 

     Period (h)        &   2.414 $\pm$ 0.078\tablenotemark{a} &  13.0 $\pm$ 1.0\tablenotemark{b}     &  2.5 $\pm$ 0.1\tablenotemark{b}      &  4.2  $\pm$  0.1\tablenotemark{b}    &  7.614 $\pm $0.178\tablenotemark{a}  &  1.369 $\pm$ 0.032\tablenotemark{a}  \\ 
%%  Parallax (mas)       &  156.25 $\pm$ 7     &                     &   136.4 $\pm$ 0.6   &  106.15$\pm$0.18    &                     &                     \\ 
%%  Distance (pc)        &   6.4 $\pm$ 0.3     &       13$\pm$ 1     &   7.33 $\pm$ 0.03   &    9.38$\pm$0.03    &      11.0$\pm$ 0.3  &        10$\pm$ 1        \\   %those with large uncertainties are photometric parallaxes.
  Discovered by        & \citet{artigau2006} &\citet{kirkpatrick2010}& \citet{dahn2002}    &  \citet{looper2007} \& \citet{metchev2008} & \citet{reid2008} & \citet{burgasser2003} \\ 

    \hline

  \end{tabular}

%  \end{sideways}
}
\tablenotetext{a}{From this work.}
\tablenotetext{b}{From \citet{metchev2015}.}
\end{table*}
%\footnotetext[1]{Determined from Spitzer data.}
\end{center}
%\end{landscape}

\begin{center}
\begin{table*}
  \caption{Journal of Observations}
   \label{obs2}
%\rotate
%\centering
\resizebox{\textwidth}{!}{%
 % \begin{sideways}
  \begin{tabular}{ cccccccc}
    \hline
     Target            &     SIMP0136        &        2M1324       &        2M1507       &        2M1821       &       2M2139        &      2M2228         \\
    \hline
    \hline
\emph{Spitzer} V1         & 2013-03-03 02:10:17 & 2013-04-02 04:44:33 & 2013-04-20 19:21:18 & 2013-06-08 20:27:01 & 2012-12-28 12:11:16 & 2012-12-18 12:24:15 \\
  Duration (hr)        &     10.04           &       54.42         &     19.99           &       16.84         &     33.45           &     5.75            \\
\emph{Spitzer} V2         & 2013-03-07 14:26:08 & 2013-04-26 04:35:30 & 2013-04-29 21:53:31 & 2013-06-15 20:21:43 & 2013-01-11 12:02:18 & 2012-12-20 17:15:18 \\
  Duration (hr)        &     10.04           &       53.71         &     19.99           &       16.84         &     33.44           &     5.75            \\
\emph{Spitzer} V3         & 2013-03-14 03:22:42 & 2013-05-27 02:49:03 & 2013-05-12 12:12:52 & 2013-06-26 21:13:47 & 2013-01-28 02:26:54 & 2012-12-24 18:09:54 \\
  Duration (hr)        &     10.04           &       53.56         &     19.99           &       16.85         &     33.44           &     5.75            \\
\emph{Spitzer} V4         & 2013-03-18 02:18:44 & 2013-06-23 10:26:47 & 2013-05-18 19:59:02 & 2013-07-01 23:08:20 & 2013-02-03 01:43:47 & 2012-12-27 13:18:29 \\
  Duration (hr)        &     10.04           &       53.56         &     19.99           &       16.85         &     33.44           &     5.75            \\
\emph{Spitzer} V5         & 2013-09-28 10:41:03 & 2014-04-18 14:43:32 & 2013-09-25 09:41:42 & 2013-11-05 02:38:36 & 2014-01-10 13:06:19 & 2013-07-20 14:41:36 \\
  Duration (hr)        &     10.05           &       56.96         &     19.98           &       16.81         &     31.15           &     7.95            \\
\emph{Spitzer} V6         & 2013-10-02 14:54:31 & 2014-05-10 16:43:12 & 2013-10-04 01:54:15 & 2013-11-11 11:57:36 & 2014-01-23 03:27:06 & 2013-07-23 15:06:46 \\
  Duration (hr)        &     10.05           &       55.27         &     19.98           &       16.81         &     31.15           &     7.95            \\
\emph{Spitzer} V7         & 2013-10-07 19:43:58 & 2014-07-16 03:11:02 & 2013-10-16 00:49:26 & 2013-11-21 12:46:43 & 2014-02-01 01:21:38 & 2013-07-27 06:29:55 \\
  Duration (hr)        &     10.05           &       55.58         &     19.98           &       16.81         &     31.15           &     8.24            \\
\emph{Spitzer} V8         & 2013-10-12 16:24:57 & 2014-08-15 09:43:18 & 2013-10-23 03:59:12 & 2013-11-27 14:17:20 & 2014-02-09 18:50:24 & 2013-07-29 03:36:11 \\
  Duration (hr)        &     10.04           &       55.84         &     19.98           &       16.81         &     31.15           &     9.94            \\
%Ch1 Exposure Time (s)  & 10.40               & 10.40               & 10.40               &  10.40              &  10.40              & 10.40               \\
%Ch2 Exposure Time (s)  & 10.40               & 10.40               & 10.40               &  10.40              &  10.40              & 10.40               \\
% didn't count the 'first' files
No. of \emph{Spitzer} Images &   21848             &   108006            &   43672             &   36641             &     70640           &    15360            \\
\emph{HST} V1             & 2013-09-28 10:44:56 &                     & 2013-04-30 10:38:04 & 2013-06-09 09:31:12 &                     & 2013-07-20 14:32:19 \\
     Overlaps with     & Spitzer V5          &                     & Spitzer V2          & Spitzer V1          &                     & Spitzer V5          \\
\emph{HST} V2             & 2013-10-07 21:10:51 &                     & 2013-05-12 16:08:48 & 2013-06-27 06:29:34 &                     & 2013-07-27 09:09:39 \\ 
     Overlaps with     & Spitzer V7          &                     & Spitzer V3          & Spitzer V3          &                     & Spitzer V7          \\
\emph{HST} Exposure Time (s)  &   112.01            &                     &   67.32             &    112.01           &                     &      223.74         \\
%   \# of Exposures    &   19x3x2 = 114      &   10x4x2 = 80       & 19x3x2 = 114        & 30x4x2 = 240        &                     &                     \\
%   Readout Mode       & SPARS25,NSAMP=6     & SPARS25,NSAMP=11    & SPARS25,NSAMP=6     &  SPARS25,NSAMP=4    &                     &                     \\
    \hline
  \end{tabular}

%  \end{sideways}
}
\end{table*}
%\footnotetext[1]{Determined from Spitzer data.}
\end{center}

%\newcolumntype{d}[3]{D{.}{\cdot}{#1} }

\begin{center}
\begin{table*}[!bht]
\small
\centering
  \caption{Phase Shifts Between \emph{Spitzer} Ch1 and Ch2 Light Curves Measured from Maxima and Minima in Rotations $2-4$ of a Visit. 
           {Visits that show strong light curve evolution and/or have incomplete phase coverage are excluded.}}
   \label{shiftmaxmin}
%\resizebox{\textwidth}{!}{
% \begin{sideways}
  \begin{tabular}{ cccccc}
    \hline

       Target       &       Visit   & \multicolumn{4}{c}{Time and Phase Shift}    \\ \cline{3-6}
                    &               &$\Delta T_{\rm{max}}$ [hr]& $\Delta\phi$ ($^{\circ}$)&$\Delta T_{\rm{min}}[hr]$& $\Delta\phi$ ($^{\circ}$) \\
  \hline
  \hline
\textbf{SIMP0136} &           1   &-0.06 $\pm$ 0.07  & -9.1 $\pm$ 11.0 & 0.04  $\pm$ 0.09  &   5.3  $\pm$  12.7 \\
                    &           2   &-0.06 $\pm$ 0.11  & -0.6 $\pm$ 15.8 & 0.07 $\pm$ 0.08  &  11.1  $\pm$  13.0 \\ 
                    &           3   & 0.09 $\pm$ 0.04  & 13.5 $\pm$  6.4 &-0.08 $\pm$ 0.07  & -11.6  $\pm$  10.7 \\
                    &           4   & 0.03 $\pm$ 0.03  &  4.3 $\pm$  4.9 &-0.15 $\pm$ 0.18  & -22.1  $\pm$  26.4 \\
                    &           6   & 0.02 $\pm$ 0.04  &  2.7 $\pm$  6.3 &-0.06 $\pm$ 0.10  &  -8.2  $\pm$  14.5 \\
                    &           7   &-0.09 $\pm$ 0.13  &-13.6 $\pm$ 19.8 &-0.21 $\pm$ 0.11  & -31.2  $\pm$  16.9 \\
  \hline
\textbf{2M1324}   &           1   &-0.59 $\pm$ 0.16  &-16.1 $\pm$  4.1 & 0.14 $\pm$ 0.12  &   3.7  $\pm$   3.3 \\
                    &           4   &-0.21 $\pm$ 0.17  & -5.8 $\pm$  4.6 & 0.25 $\pm$ 0.17  &   6.8  $\pm$   4.5 \\
                    &           5   &-0.01 $\pm$ 0.04  & -0.2 $\pm$  1.2 &-0.14 $\pm$ 0.17  &  -3.8  $\pm$   4.7 \\
  \hline
\textbf{2M1507}   &           3   &-0.09 $\pm$ 0.11  &-13.4 $\pm$ 15.8 & 0.08 $\pm$ 0.11  &  11.0  $\pm$  16.2 \\
  \hline
%%\textbf{2M1821}   &           1   &-0.031 $\pm$ 0.264  & -2.8 $\pm$ 23.8 & 0.005 $\pm$ 0.127  &   0.5  $\pm$  11.4 \\
\textbf{2M1821}   &           2   &-0.01 $\pm$ 0.20  & -1.1 $\pm$ 17.7 & 0.20 $\pm$ 0.17  &  18.4  $\pm$  16.7 \\ 
         %%         &           3   &-0.089 & -7.97& & \\
                    &           4   & 0.013 $\pm$ 0.23  &  1.2 $\pm$ 21.0 & 0.05 $\pm$ 0.18  &   4.6  $\pm$  16.4 \\
                    &           6   & 0.18 $\pm$ 0.12  & 15.8 $\pm$ 10.8 & 0.43 $\pm$ 0.20  &  38.3  $\pm$  18.2 \\
                    &           8   & 0.35 $\pm$ 0.15  & 31.3 $\pm$ 13.9 & 0.04 $\pm$ 0.14  &   3.7  $\pm$  13.0 \\
  \hline
%%%\textbf{2M2139}   &           1   &-0.464 $\pm$ 0.257  &-21.94 $\pm$ 12.15 & 0.475 $\pm$ 0.053  &  22.46  $\pm$   2.51 \\
\textbf{2M2139}   &           2   & 0.21 $\pm$ 0.08  &  9.8 $\pm$  4.0 &-0.02 $\pm$ 0.08  &  -1.0  $\pm$   3.9 \\ 
                    &           3   &-0.17 $\pm$ 0.05  &  7.9 $\pm$  2.3 & 0.19 $\pm$ 0.09  &   9.0  $\pm$   4.2 \\
                    &           4   &-0.07 $\pm$ 0.05  & -3.3 $\pm$  2.2 & 0.35 $\pm$ 0.13  &  16.3  $\pm$   6.3 \\
                    &           5   & 0.02 $\pm$ 0.04  &  1.1 $\pm$  2.1 & 0.35 $\pm$ 0.13  &  16.5  $\pm$   6.3 \\
                    &           7   & 0.43 $\pm$ 0.15  & 20.4 $\pm$  7.0 &-0.08 $\pm$ 0.04  &  -3.6  $\pm$   1.7 \\

        %%          &           6   & 0.258 & 12.19& 0.545 &  25.77  \\

    \hline
  \end{tabular}
\end{table*}
\end{center}

\begin{center}
\begin{table*}[!bht]
%\scriptsize
\small
\centering
  \caption{Phase Shifts Between \emph{Spitzer} Ch1 and Ch2 Light Curves Measured from Cross-correlation.
           {Visits that show strong light curve evolution are excluded.}}
   \label{shiftcross}
%\resizebox{\textwidth}{!}{
% \begin{sideways}
  \begin{tabular}{ ccccccc}
    \hline

       Target       & $P$\tablenotemark{a} (hr) &   Visit   & $P_{\rm{Visit}}$\tablenotemark{b} (hr) & \multicolumn{2}{c}{Phase Shift}    \\ \cline{5-6}
                    &                          &           &                                        & $\Delta\phi_{\rm{R2-R3}}$ ($^{\circ}$)& $\Delta\phi_{\rm{R3-R4}}$ ($^{\circ}$) \\
  \hline
  \hline
\textbf{SIMP0136} &  2.414                   &      1   & 2.628 & -0.4 $\pm$ 4.8&    0.1 $\pm$ 4.6  \\
                    &                          &      2   & 2.623 &  7.3 $\pm$ 7.1&   -7.5 $\pm$ 6.6  \\ 
                    &                          &      3   & 2.617 & -5.7 $\pm$ 4.0&  -0.3 $\pm$ 4.0  \\
%                    &                          &      4   & 2.610 &  3.5 $\pm$ 4.0& -21.7 $\pm$ 4.8 \\
                    &                          &      6   & 2.611 & 10.7 $\pm$ 6.1& -14.8 $\pm$ 7.2 \\
                    &                          &      7   & 2.531 & 12.8 $\pm$ 6.8& -16.9 $\pm$ 7.0 \\
  \hline
\textbf{2M1324}   & 13.0                     &      1   &12.339 & 13.8 $\pm$ 27.5& -11.8 $\pm$ 27.5 \\
                    &                          &      4   &12.346 & -2.6 $\pm$ 1.6 &   3.8 $\pm$ 1.8 \\
  \hline
\textbf{2M1821}   & 4.2                      &      1   & 4.177 & -1.8 $\pm$ 9.5 & -5.9 $\pm$ 9.1  \\
%                    &                          &      2   & 4.481 &-18.49& -11.69  \\ 
%                    &                          &      3   & 4.426 &-18.70&  12.13  \\ 
                    &                          &      4   & 4.262 & -8.8 $\pm$ 5.6 &   1.5 $\pm$ 6.7  \\
                    &                          &      5   & 4.289 &  4.1 $\pm$ 5.8 &  -3.9 $\pm$ 5.4 \\
                    &                          &      8   & 4.439 & -2.9 $\pm$ 6.0 &  -8.5 $\pm$ 6.5 \\
  \hline
\textbf{2M2139}   &  7.614                   &      5   & 7.790 & -1.0 $\pm$ 9.0 &   0.2 $\pm$ 2.2 \\
                    &                          &      6   & 7.775 &-16.3 $\pm$ 21.5 &  18.1 $\pm$ 22.3 \\
                    &                          &      7   & 7.654 & -2.8 $\pm$ 1.6 &   3.2 $\pm$ 1.7 \\ 
  \hline
\textbf{2M2228}   &   1.369                  &      3   & 1.516 & 8.6 $\pm$ 10.6 &  -5.6 $\pm$ 9.2 \\
    \hline

  \end{tabular}
%\end{sideways} %}
\tablenotetext{a}{Rotation period measured from cross-correlating all 8 visits of data.}
\tablenotetext{b}{Rotation period measured in the specific visit from cross-correlating the Ch1 light curves of Rotation 2 and Rotation 4. }
\end{table*}
\end{center}

\begin{center}
\begin{table*}[!bht]
  \caption{Phase Shifts Between \emph{HST} $J$-band and \emph{Spitzer} Light Curves Measured from Fourier Fits. {\emph{Spitzer} Visit 7 of SIMP0136 and Visit 1 of 2M1507 are excluded due to substantially different light curve shapes between different wavelength bands.}}
   \label{shift1}
\resizebox{\textwidth}{!}{
  \begin{tabular}{ ccccc}
    \hline

     Target            &   Phase Shift Between                          & 1st Fourier Component   & 2nd Fourier Component     \\
                       &                                                &$\Delta\phi$ ($^{\circ}$) &$\Delta\phi$ ($^{\circ}$)   \\
    \hline
    \hline
\textbf{SIMP0136}        &\emph{HST} V1 and \emph{Spitzer} V5 Ch1         &      33.4 $\pm$ 3.9   &       16.0 $\pm$ 34.4         \\
%%                       &\emph{HST} V2 and \emph{Spitzer} V7 Ch1         &     -12.1 $\pm$ 5.1   &       98.3 $\pm$ 57.3        \\ 
    \hline
%%\textbf{2M1507}          &\emph{HST} V1 and \emph{Spitzer} V1 Ch2         &     197.7 $\pm$ 9.5   &       80.7 $\pm$ 55.0        \\
%%                       &\emph{HST} V1 and \emph{Spitzer} V1 Ch1         &     207.7 $\pm$ 11.3  &      110.4 $\pm$ 49.1        \\
\textbf{2M1507}      &\emph{HST} V2 and \emph{Spitzer} V3 Ch1         &     171.1 $\pm$ 5.3   &      245.8 $\pm$ 39.0        \\
    \hline
\textbf{2M1821}          &\emph{HST} V1 and \emph{Spitzer} V1 Ch1         &     167.6 $\pm$ 2.2   &      147.8 $\pm$ 41.7        \\
                       &\emph{HST} V2 and \emph{Spitzer} V3 Ch2         &     195.3 $\pm$ 3.4   &      237.7 $\pm$ 10.7        \\
    \hline
\textbf{2M2228}          &\emph{HST} V1 and \emph{Spitzer} V5 Ch1         &     158.9 $\pm$ 4.6   &      125.5 $\pm$ 38.1        \\
                       &\emph{HST} V1 and \emph{Spitzer} V5 Ch2         &     156.5 $\pm$ 9.5   &       38.5 $\pm$ 97.1        \\
%%                       &\emph{HST} V2 and \emph{Spitzer} V7 Ch1 (pt. 1) &     149.8 $\pm$ 8.5   &      -41.2 $\pm$ 34.5        \\
%%                       &\emph{HST} V2 and \emph{Spitzer} V7 Ch1 (pt. 2) &     184.4 $\pm$ 4.8   &       61.3 $\pm$ 33.2        \\
                       &\emph{HST} V2 and \emph{Spitzer} V7 Ch1         &     167.4 $\pm$ 3.8   &      -11.7 $\pm$ 30.8        \\

    \hline
  \end{tabular}

}
\end{table*}
\end{center}

%%\begin{center}
%%\begin{table}
%%  \caption{Best-fit Atmospheric Model Parameters for the Targets.}
%%   \label{modelparam}
%%\resizebox{\textwidth}{!}{
%%  \begin{tabular}{ cccc}
%%    \hline
%%
%%                                &     \multicolumn{3}{c}{Best-fit Parameters} \\ \cline{2-4}
%%     Target                     &    \Teff (K)       &  \logg        &   $f_{\rm{sed}}$     \\
%%    \hline
%%    \hline
%%2M1507 (L5) and 2M1821 (L5)     &     1700           &     3.0       &     3       \\
%%2M2139 (T2  ) and SIMP0136 (T2) &     1400           &     2.5       &     5       \\
%%2M1324 (T2)\tablenotemark{a}    &     1400           &     2.5       &     5       \\
%%2M2228 (T6)                   &      950           &     2.5       &     5       \\
%%
%%    \hline
%%  \end{tabular}
%%}
%%
%%\tablenotetext{a}{Due to the absence of high-precision \emph{HST}/WFC3 spectrum of 2M1324, we use the best-fit parameters from fits of SIMP0136, which has the same 
%%                  spectral type as 2M1324.}
%%\end{table}
%%\end{center}

\begin{center}
\begin{table*}
  \caption{Best-fit Atmospheric Model Parameters for the Targets.}
   \label{modelparam}
\resizebox{\textwidth}{!}{
  \begin{tabular}{ccccc}
    \hline
%                                &     \multicolumn{3}{c}{Best-fit Parameters} \\ \cline{2-4}
     Target          &                           &  2M1507                        & 2M2139, SIMP0136            & 2M2228       \\
                     &                           &    \&  2M1821                  & \&  2M1324\tablenotemark{a} &               \\ \cline{1-1}\cline{3-5}
   % \hline
   Spectral Type     &                           &          L5                    &             T2              &   T6          \\
    \hline
    \hline
Best-fit Model Parameters  &     \Teff (K)              &               1700           &     1400      &     950     \\
                     &     \logg                  &               5.0            &     5.0       &     4.5     \\
                     &     $f_{\rm{sed}}$         &               3              &       5       &     5       \\
    \hline
Model Pressure (bar) & $1.12-1.17~\mu$m (H$_2$O and alkali)&    6.55             &    7.60       &    6.99     \\
                     & $1.21-1.32~\mu$m (narrow $J$)&           6.50             &    8.12       &    7.59     \\
                     & 2MASS $J$-band             &             6.48             &    7.29       &    6.93     \\
                     & $1.35-1.43~\mu$m (H$_2$O)  &             4.33             &    4.13       &    1.91     \\
                     & $1.54-1.60~\mu$m (narrow $H$)&           5.94             &    7.08       &    6.04     \\
                     & $1.62-1.69~\mu$m (CH$_4$ and H$_2$O) &   6.02             &    7.18       &    4.26     \\
                     & \emph{Spitzer} Ch1         &             2.75             &    0.37       &    0.24     \\
                     & \emph{Spitzer} Ch2         &             1.59             &    1.00       &    0.48     \\
    \hline
  \end{tabular}
}

\tablenotetext{a}{Due to the absence of high-precision \emph{HST}/WFC3 spectrum of 2M1324, we use the best-fit parameters from fits of SIMP0136, which has the same 
                  spectral type as 2M1324.}
\end{table*}
\end{center}

%%Model Pressure (bar) & $1.12-1.17~\mu$m (H$_2$O and alkali)&    4.01             &    3.78       &    6.99     \\
%%                     & $1.21-1.32~\mu$m (narrow $J$)&           4.49             &    4.13       &    7.59     \\
%%                     & 2MASS $J$-band             &             4.09             &    3.87       &    6.93     \\
%%                     & $1.35-1.43~\mu$m (H$_2$O)  &             1.96             &    1.96       &    1.91     \\
%%                     & $1.54-1.60~\mu$m (narrow $H$)&           3.03             &    3.12       &    6.04     \\
%%                     & $1.62-1.69~\mu$m (CH$_4$ and H$_2$O) &   3.03             &    3.45       &    4.26     \\
%%                     & \emph{Spitzer} Ch1         &             1.43             &    0.29       &    0.24     \\
%%                     & \emph{Spitzer} Ch2         &             0.77             &    1.00       &    0.48     \\

%%\clearpage

%%%%%%%%%%%%%%%%%%%%%%%%%
%Figures
%%%%%%%%%%%%%%%%%%%%%%%%%

%%residuals
\begin{figure*}[ht]
  \begin{center}
    \includegraphics[scale=0.7]{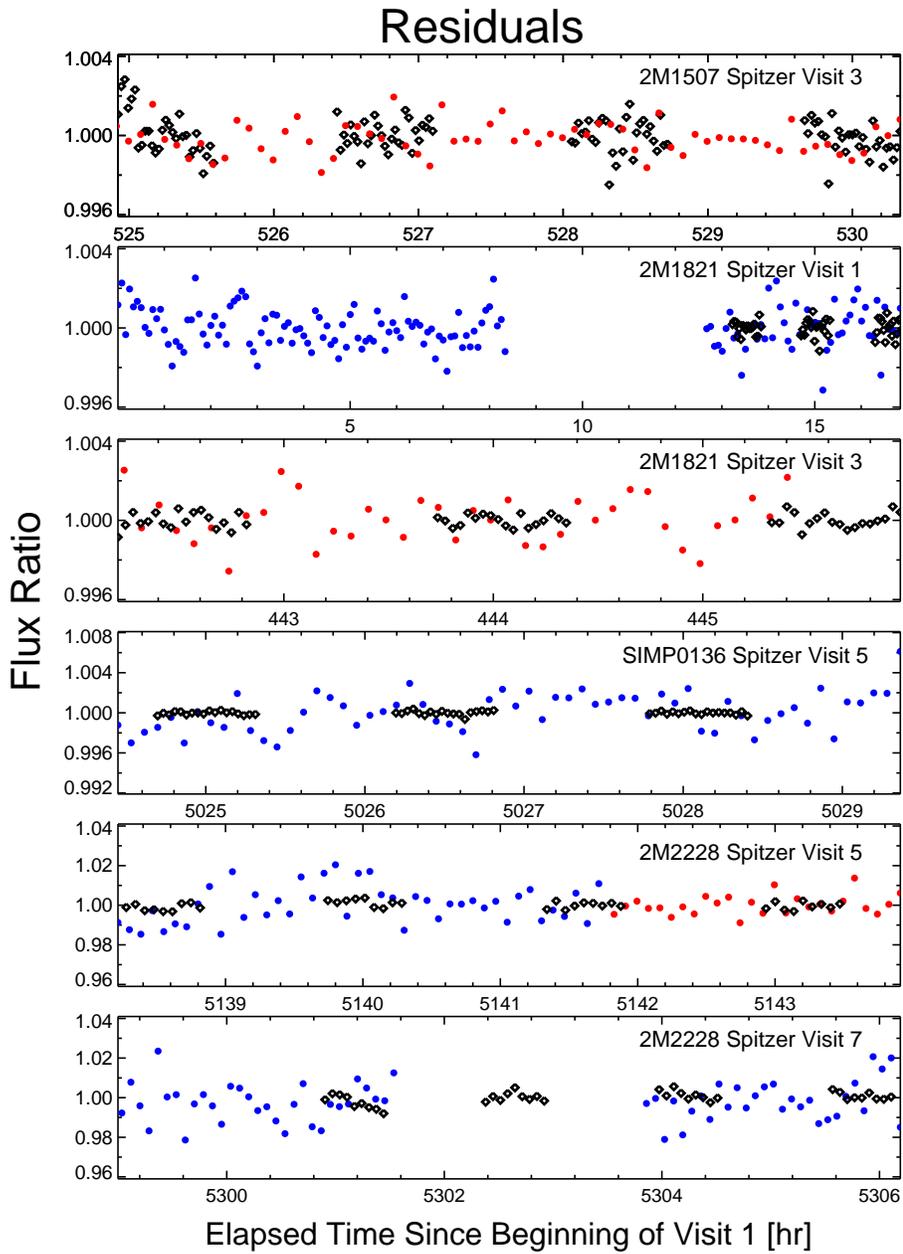}
     \vspace*{7mm}
       \caption{ Light curve residuals from the Fourier fits.                
               The blue and red dots represent residuals for the \emph{Spitzer} Ch1 and Ch2 light curves, respectively. 
	       The black open diamonds are \emph{HST} $J$-band residuals.} 
           \label{residuals}
              \end{center}
    \end{figure*}

%%autocorrelation

\begin{figure*}[ht]
  \begin{center}
      \includegraphics[scale=0.70, angle=90]{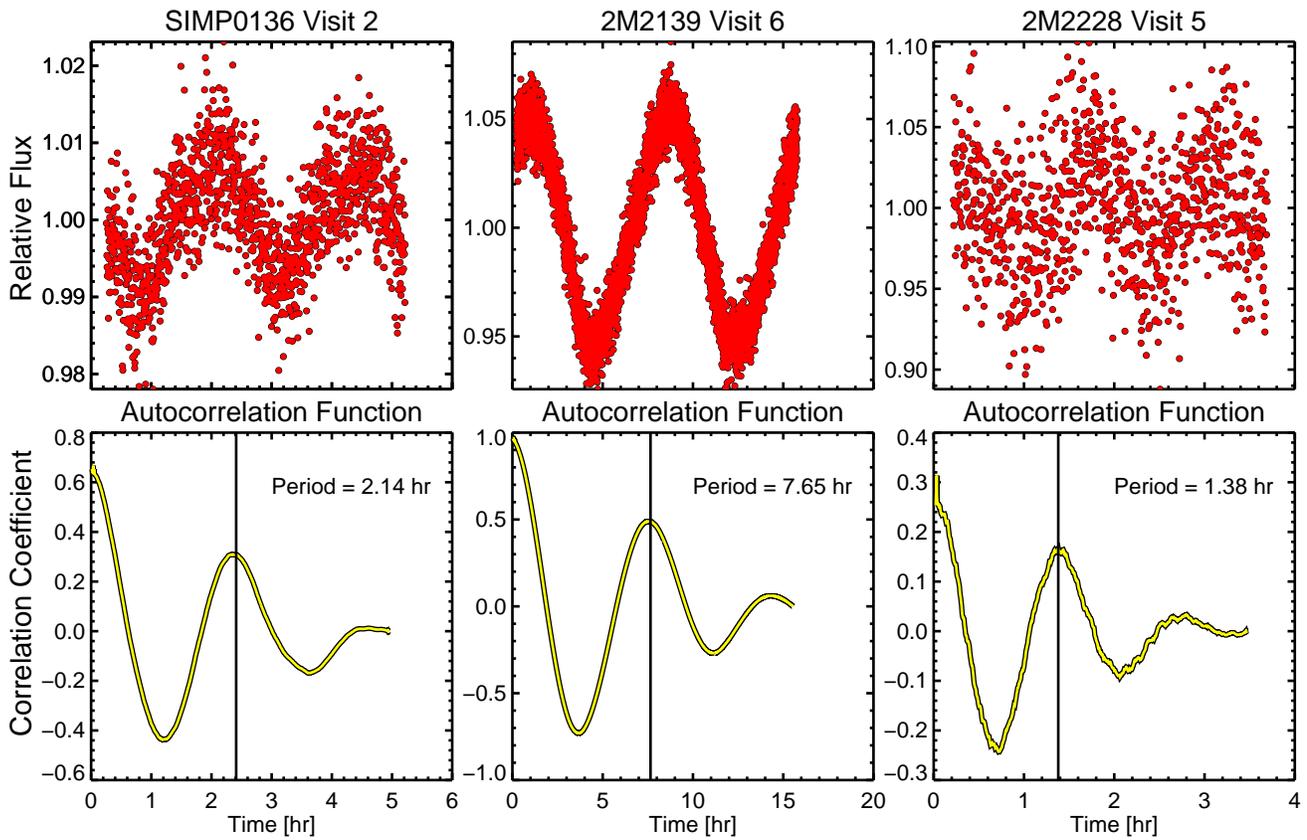}
        \caption{ Examples of the autocorrelation analysis for SIMP0136, 2M2139, and 2M2228. The {top panels} show the unbinned 
	  data, and the {bottom panels} show the autocorrelation functions and the identified peaks.}
	\label{autocorr}
\end{center}
\end{figure*}

\begin{figure*}
  \begin{center}
    \includegraphics[scale=0.8]{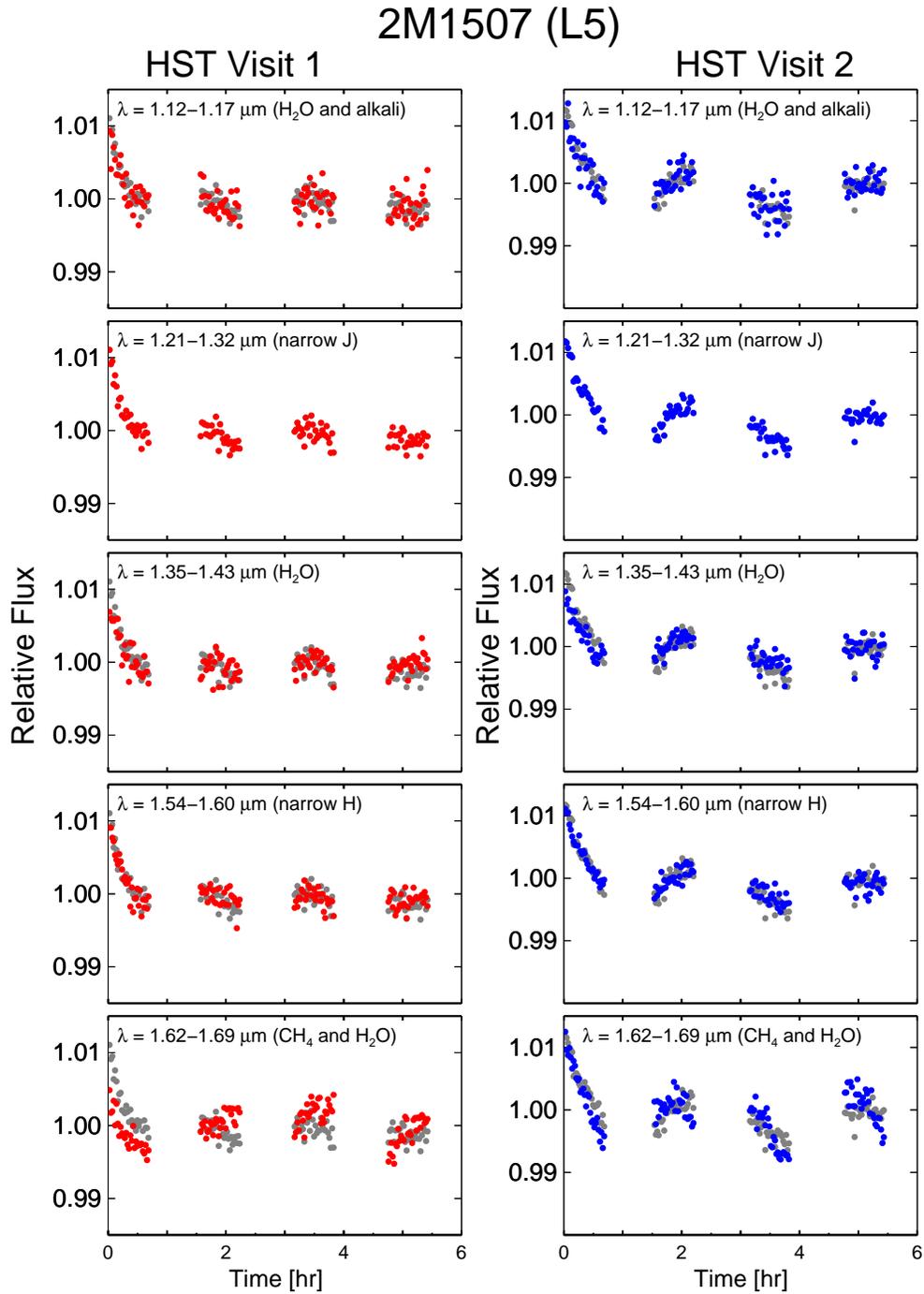}
       \caption{\emph{HST} narrow-band light curves of 2M1507. {The dark gray points are the $J$-band light curves and show comparison with the light curves of other narrow bands.} }
           \label{narrow1507}
              \end{center}
    \end{figure*}

\begin{figure*}
  \begin{center}
    \includegraphics[scale=0.8]{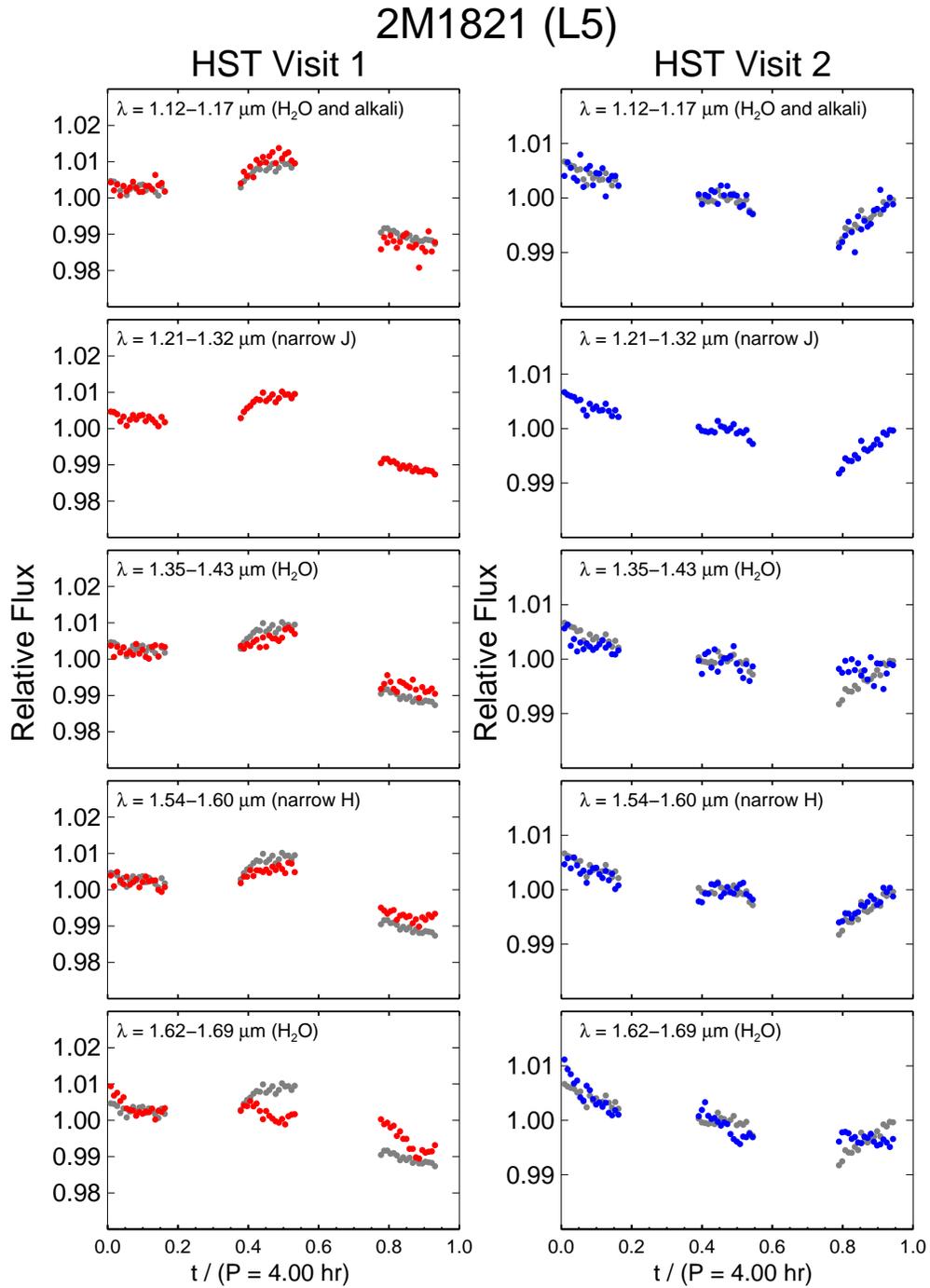}
       \caption{Phase-folded \emph{HST} narrow-band light curves of 2M1821. {The dark gray points are the $J$-band light curves and show comparison with the light curves of other narrow bands.}}
           \label{narrow1821}
              \end{center}
    \end{figure*}

\clearpage

\begin{figure*}
  \begin{center}
    \includegraphics[scale=0.8]{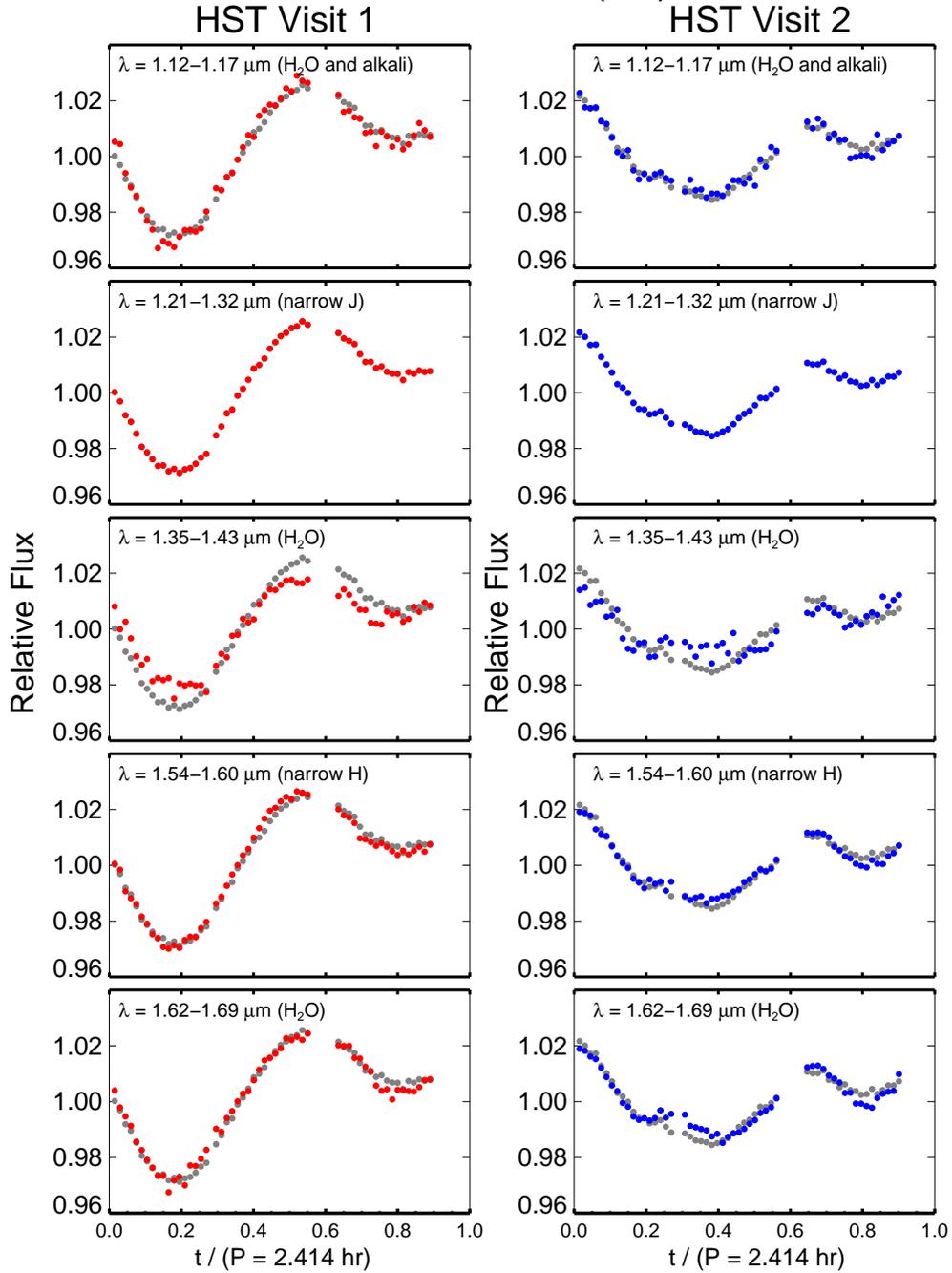}
       \caption{Phase-folded \emph{HST} narrow-band light curves of SIMP0136. {The dark gray points are the $J$-band light curves and show comparison with the light curves of other narrow bands.}
                No detectable phase shift was found between the five narrow bands.}
           \label{narrow0136}
              \end{center}
    \end{figure*}

\begin{figure*}
  \begin{center}
    \includegraphics[scale=0.8]{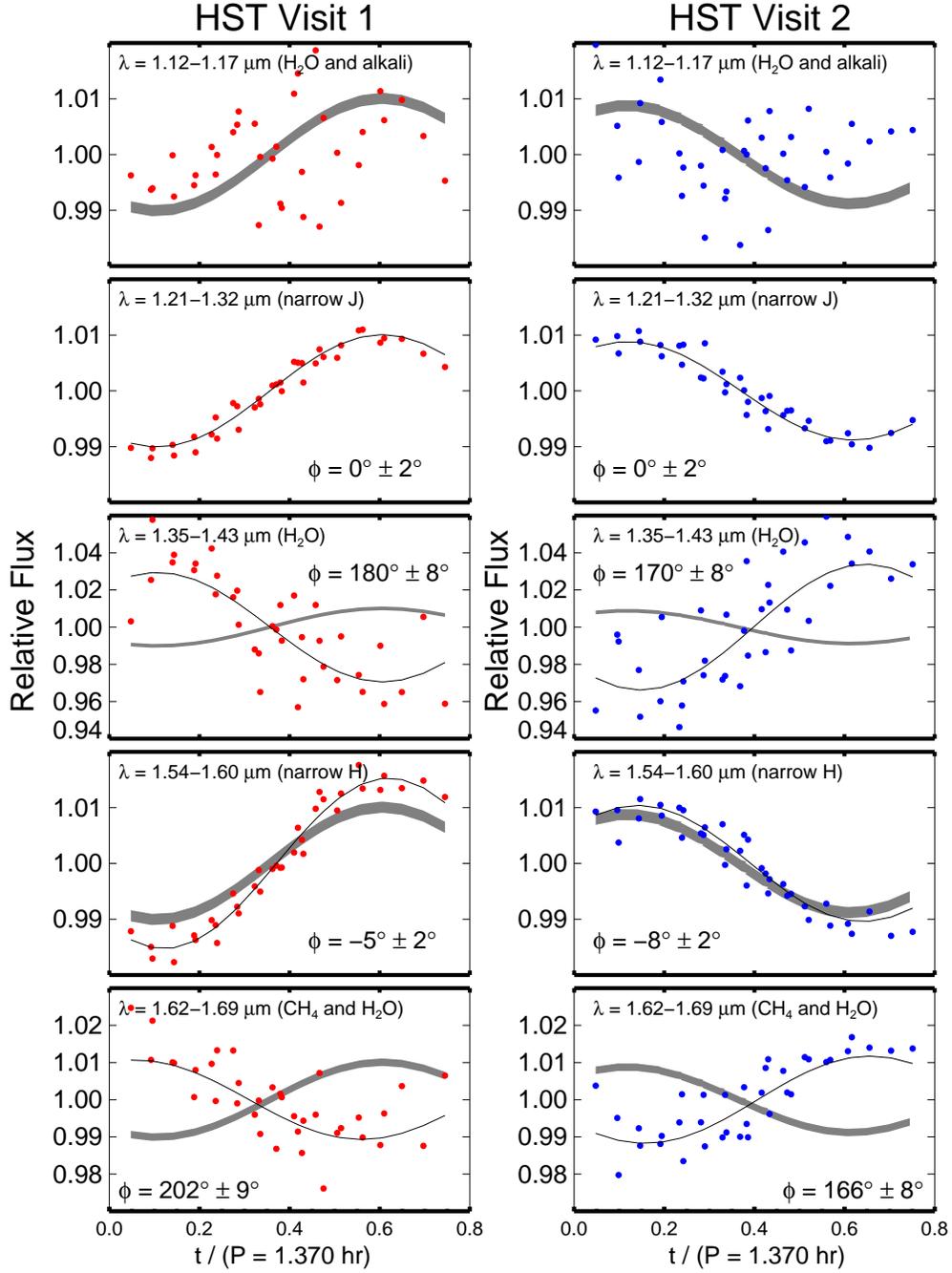}
       \caption{Phase-folded \emph{HST} narrow-band light curves of 2M2228. 
               {The solid lines are the best-fit sine-wave models, and the dark gray bands are the best-fit models to the $J$-band light curves, with the widths of the bands 
	       representing the scatter in the data.}
                We detected phase shifts between four narrow bands, and the results are consistent with that measured by \citet{buenzli2012}.
                }
           \label{narrow2228}
              \end{center}
    \end{figure*}

%%correlations
%\begin{figure}[ht]
%  \begin{center}
%    \includegraphics[scale=0.75, angle=90]{shft.rotation.ps}
%       \caption{ Phase shift between \emph{HST} $J$-band and \emph{Spitzer} light curves is plotted as a function of rotation period. 
%                We found no obvious correlation between phase shift and phase shift.
%                }
%           \label{shftrotation}
%              \end{center}
%    \end{figure}
%
%
%\begin{figure}[ht]
%  \begin{center}
%    \includegraphics[scale=0.75, angle=90]{shft.sptype.ps}
%       \caption{ Phase shift between \emph{HST} $J$-band and \emph{Spitzer} light curves is plotted as a function of spectral type. 
%       The phase shifts of 2M1507 (L5), 
%         2M1821 (L5), and 2M2228 (T6) are centered around 180$^\circ$, while SIMP0136 (T2) shows a phase shift about 30$^\circ$. }
%           \label{shftsptype}
%              \end{center}
%    \end{figure}

\begin{figure*}
  \begin{center}
    \includegraphics[scale=0.7, angle=0]{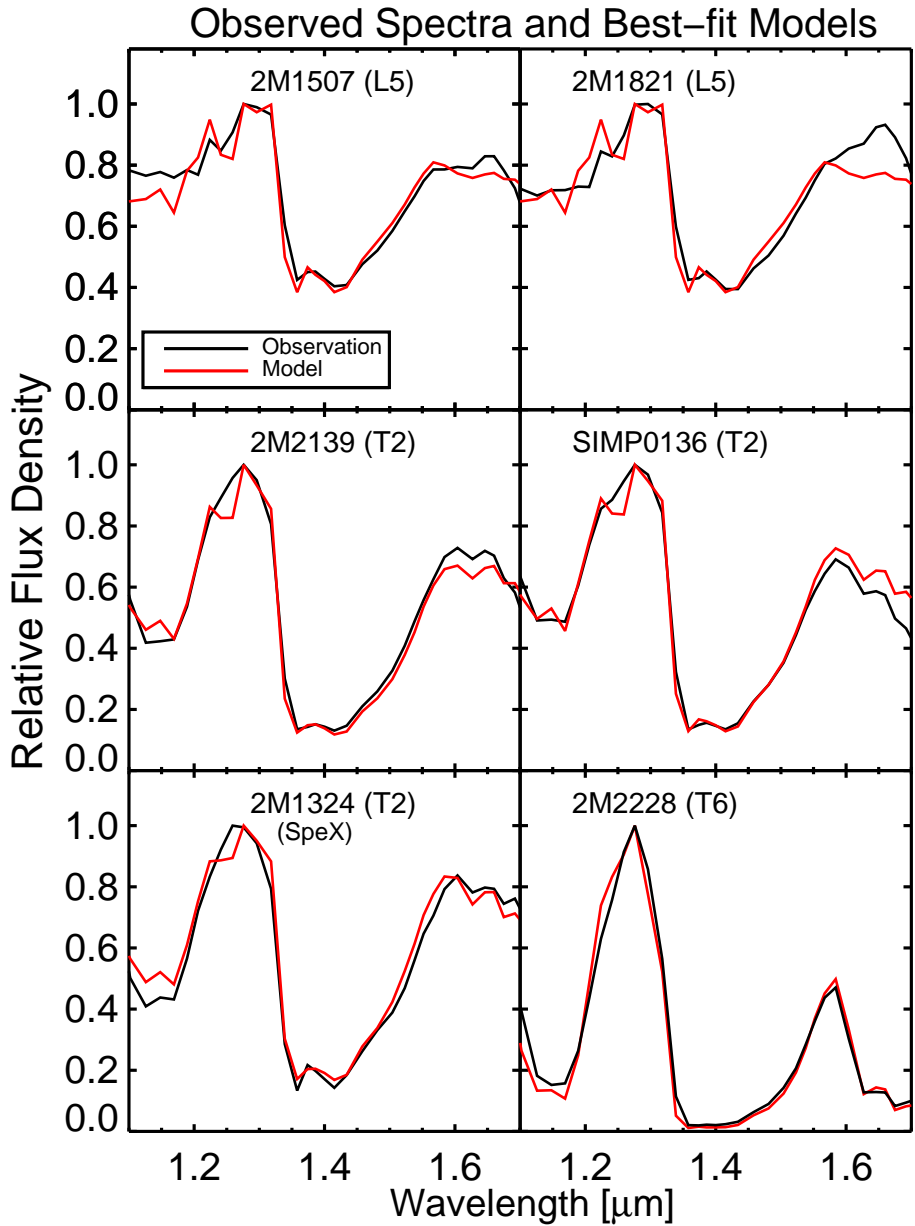}
       \vspace*{5mm}
       \caption{\emph{HST}/WFC3 grism spectra of five dwarfs and their respective best-fit models. IRTF/SpeX spectrum of 2M1324 is also
               presented and we overplot the best-fit model from the other T2 dwarfs.
               }
           \label{modelfits}
              \end{center}
    \end{figure*}

\begin{figure*}
  \begin{center}
    \includegraphics[scale=0.70, angle=0]{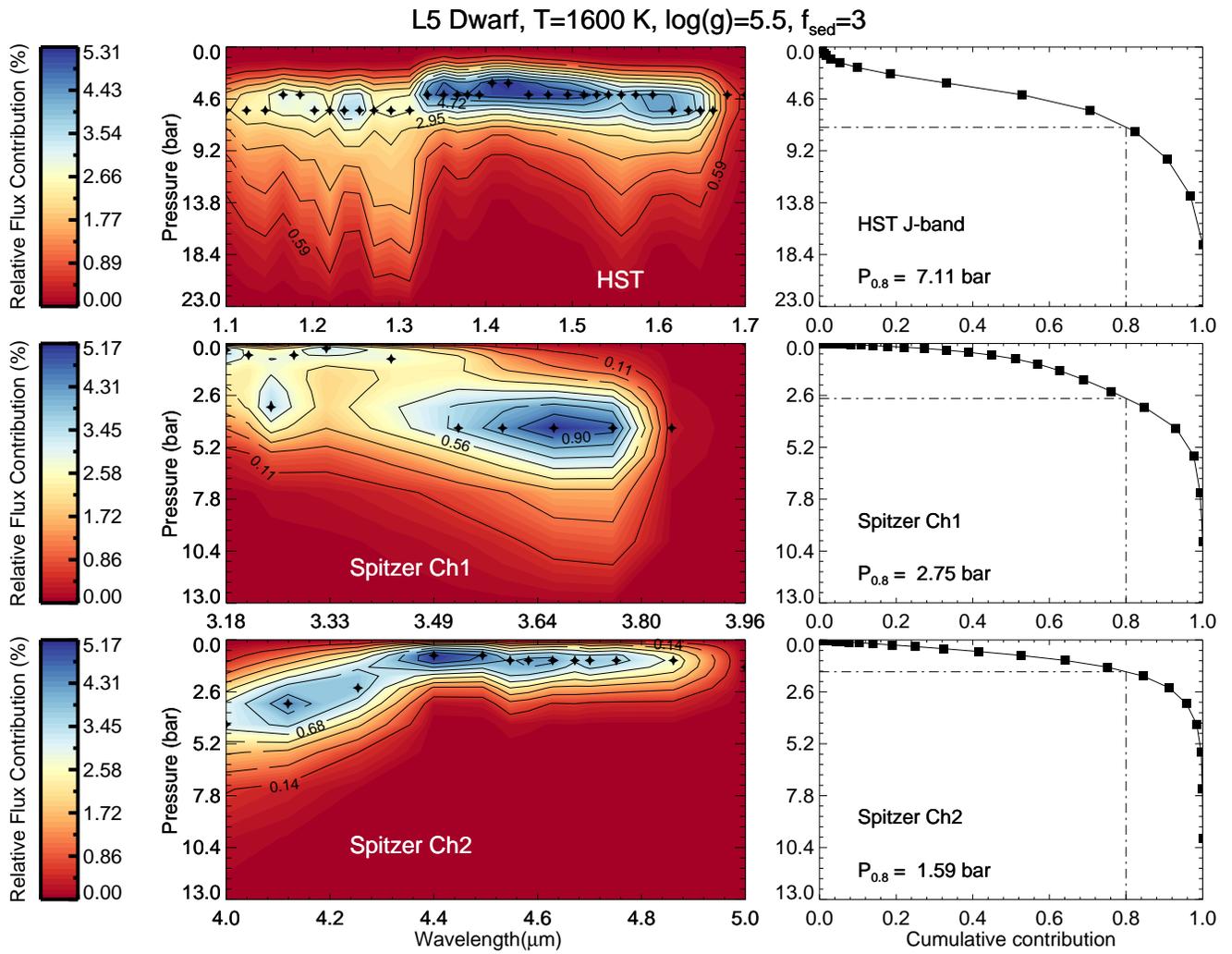}
       \caption{Relative flux contributions and cumulative contributions for the best-fit model of the two L5 targets, 2M1507 and 2M1821. 
                The model is for \Teff\ of 1600 K, \logg\ of 5.5, and $f_{\rm{sed}}$ of 3. See discussion in \S 6.1. 
               }
           \label{contrifunc1700}
              \end{center}
    \end{figure*}

\begin{figure*}
  \begin{center}
    \includegraphics[scale=0.70, angle=0]{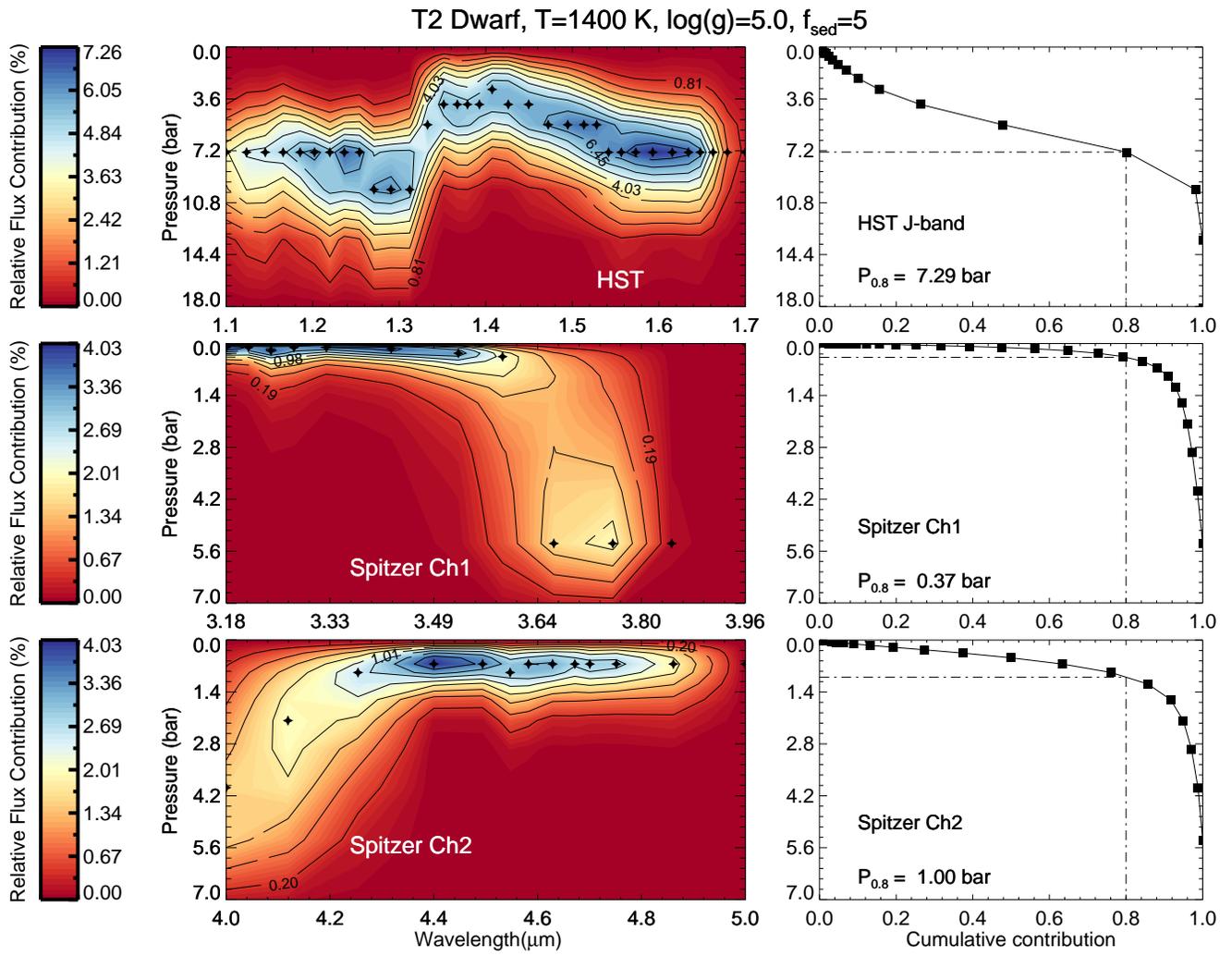}
       \caption{Relative flux contributions and cumulative contributions for the best-fit model of the T2 dwarf SIMP0136.  
                The model is for \Teff\ of 1400 K, \logg\ of 4.5, and $f_{\rm{sed}}$ of 5. See discussion in \S 6.1. 
                 }
           \label{contrifunc1400}
              \end{center}
    \end{figure*}

%\begin{figure}[ht]
%  \begin{center}
%    \includegraphics[scale=0.65, angle=0]{all_t1400g300f5.ps}
%       \caption{Relative flux contributions and cumulative contributions for the best-fit model of the T2 dwarf 2M2139.  
%                The model is for \Teff\ of 1400 K, \logg\ of 4.5, and $f_{\rm{sed}}$ of 5. See discussion in \S 6.1. 
%                 }
%           \label{contrifunc1400}
%              \end{center}
%    \end{figure}
%
%\clearpage
%
%\begin{figure}[ht]
%  \begin{center}
%    \includegraphics[scale=0.65, angle=0]{all_t1300g300f3.ps}
%       \caption{Relative flux contributions and cumulative contributions for the best-fit model of the T2 dwarf 2M1324.  
%                The model is for \Teff\ of 1300 K, \logg\ of 4.5, and $f_{\rm{sed}}$ of 3. See discussion in \S 6.1. 
%                 }
%           \label{contrifunc1400}
%              \end{center}
%    \end{figure}

\begin{figure*}
  \begin{center}
    \includegraphics[scale=0.70, angle=0]{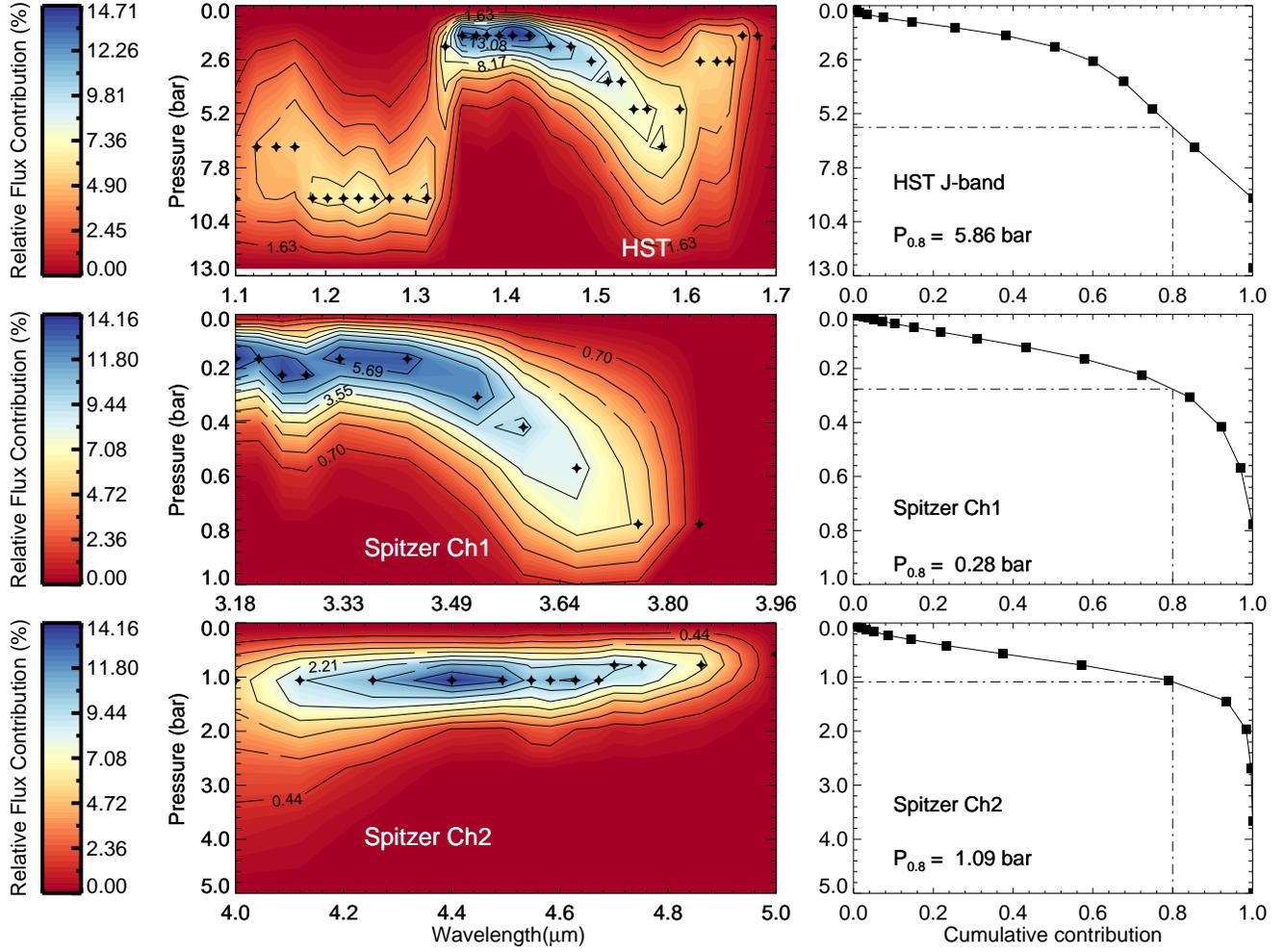}
       \caption{Relative flux contributions and cumulative contributions for the best-fit model of the T6 dwarf 2M2228.  
                The model is for \Teff\ of 950 K, \logg\ of 4.5, and $f_{\rm{sed}}$ of 5. See discussion in \S 6.1. 
               }
           \label{contrifunc950}
              \end{center}
    \end{figure*}

\begin{figure*}
  \begin{center}
    \includegraphics[scale=0.11]{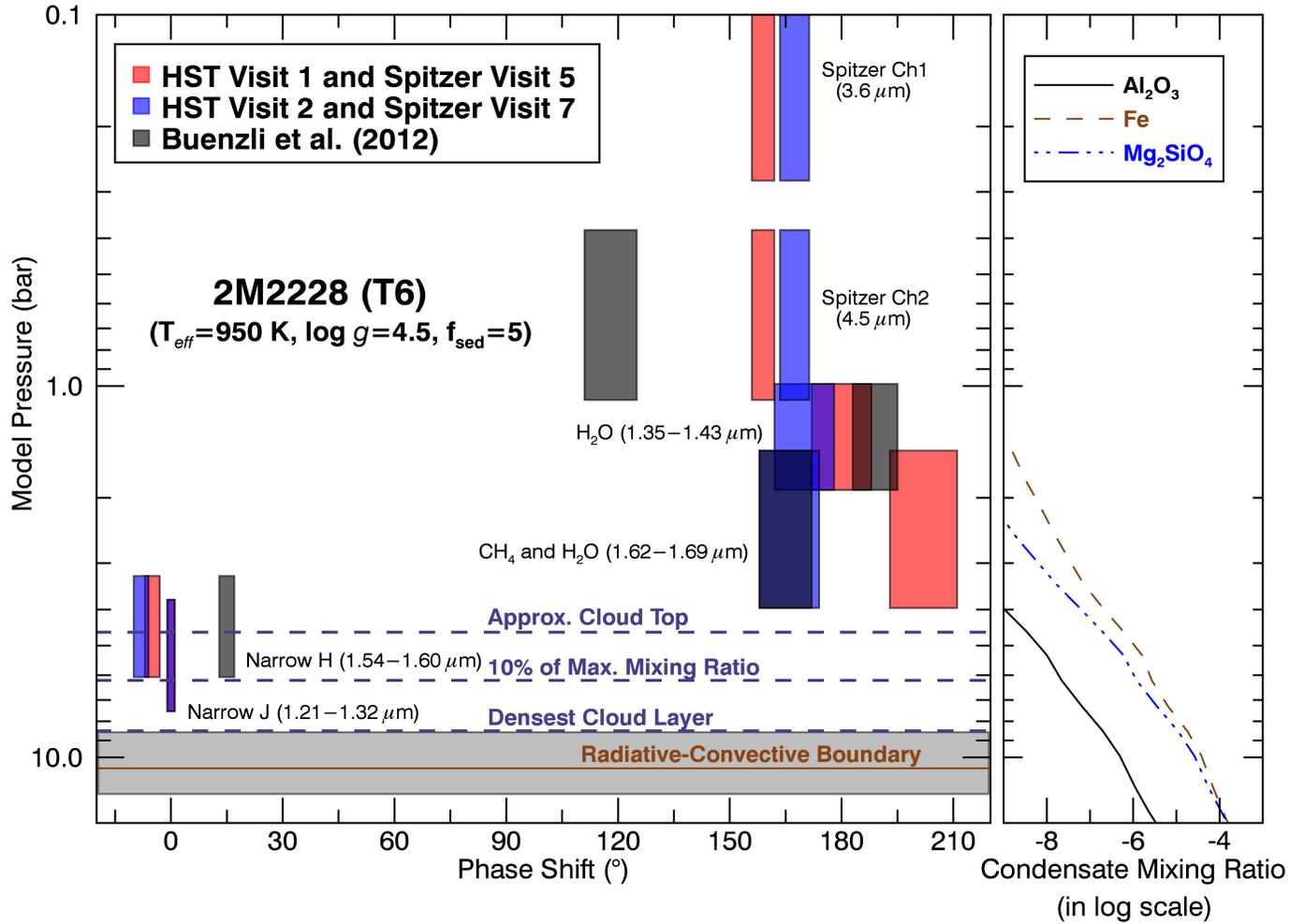}
       \caption{\emph{Left panel:} phase shifts between light curves of six spectral bands of 2M2228 (T6) plotted as a function 
                     of characteristic pressure levels probed by the bandpasses. The phase shifts are all with respect to that of 
                     the narrow \emph{J} band. The size of a rectangle in the horizontal direction marks the 1-$\sigma$ range of the phase shift
                     for a narrow bandpass, and the size of a rectangle in the vertical direction marks the pressure region in the model
                     where between 20\% and 80\% of the total flux comes from for a narrow bandpass. 
                     The solid horizontal line in brown is the radiative-convective
                     boudary in the atmospheric model, and the gray band marks the size of the particular model pressure grid. 
                     The three dashed horizontal lines represent the pressure levels where the 
                     condensate volume mixing ratio is at 100\%, 10\%, and 1\% of the maxmimum value, respectively.
                 \emph{Right panel:} the condensate volume mixing ratio in the model plotted as a function of model pressure.
               }
           \label{shftpressure2228}
              \end{center}
    \end{figure*}

\begin{figure*}
  \begin{center}
    \includegraphics[scale=0.11]{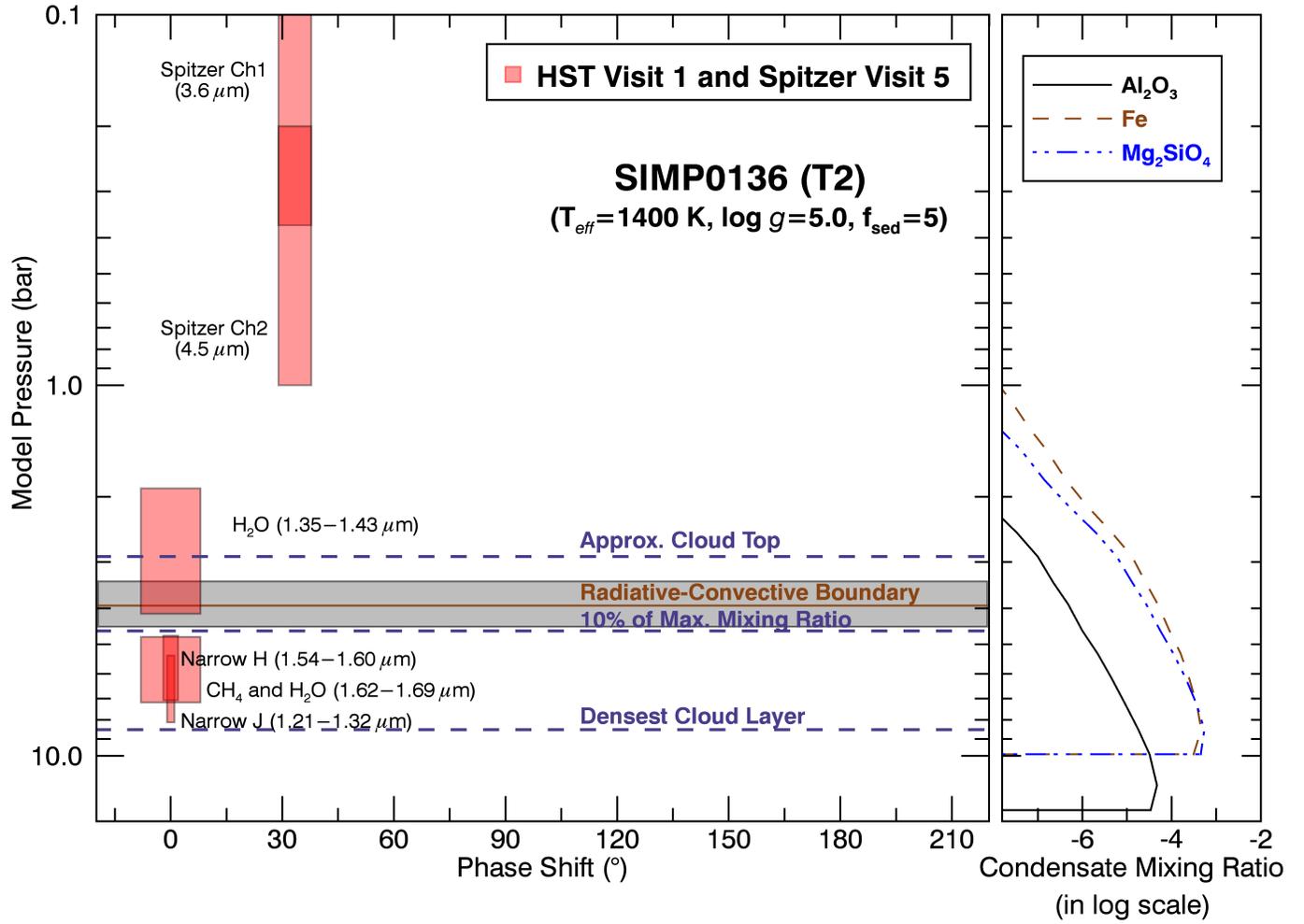}
       \caption{The same as Figure~\ref{shftpressure2228} except for SIMP0136 (T2).}
           \label{shftpressure0136}
              \end{center}
    \end{figure*}

\begin{figure*}
  \begin{center}
    \includegraphics[scale=0.11]{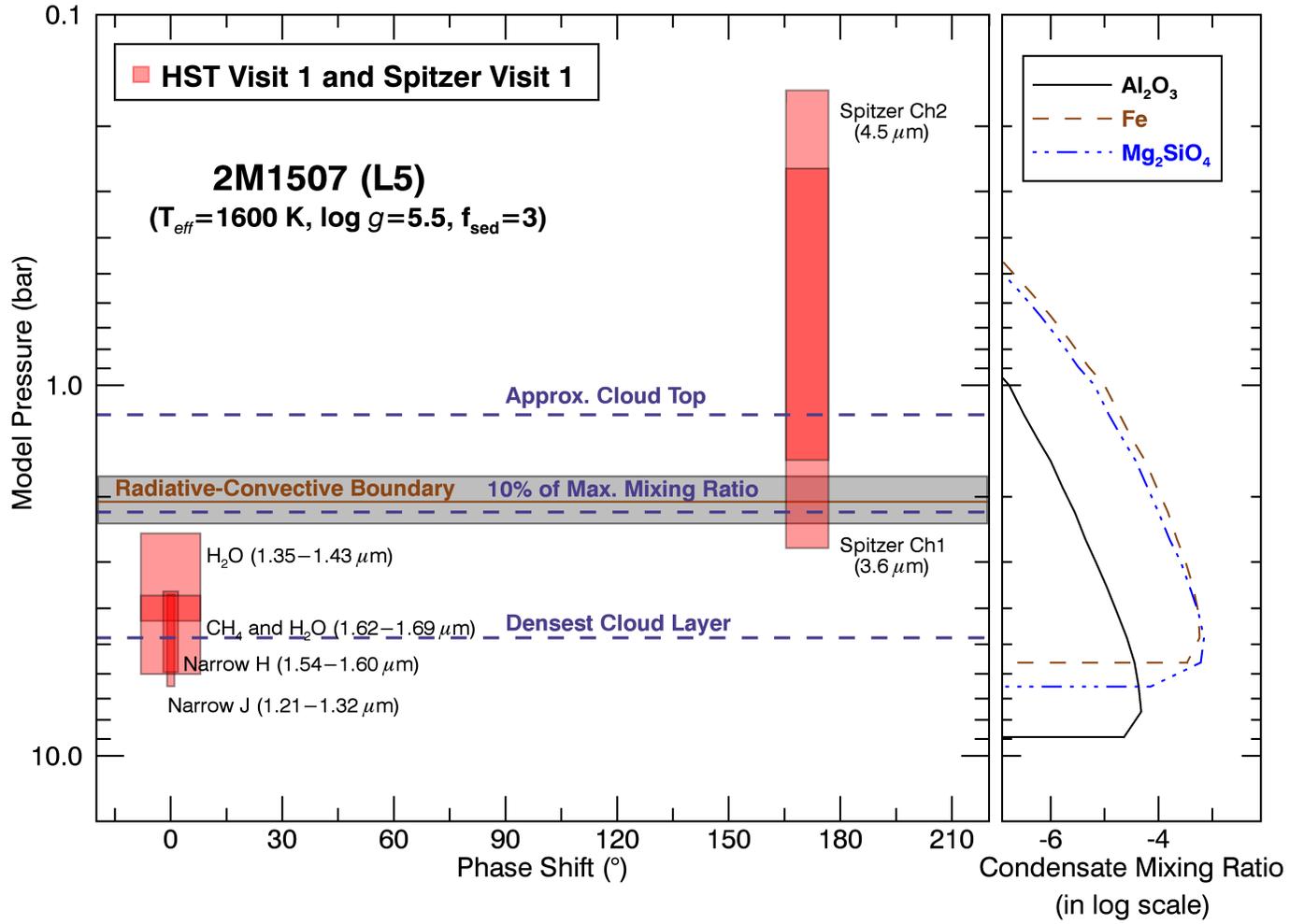}
       \caption{The same as Figure~\ref{shftpressure2228} except for 2M1507 (L5).}
           \label{shftpressure1507}
              \end{center}
    \end{figure*}

\begin{figure*}
  \begin{center}
    \includegraphics[scale=0.11]{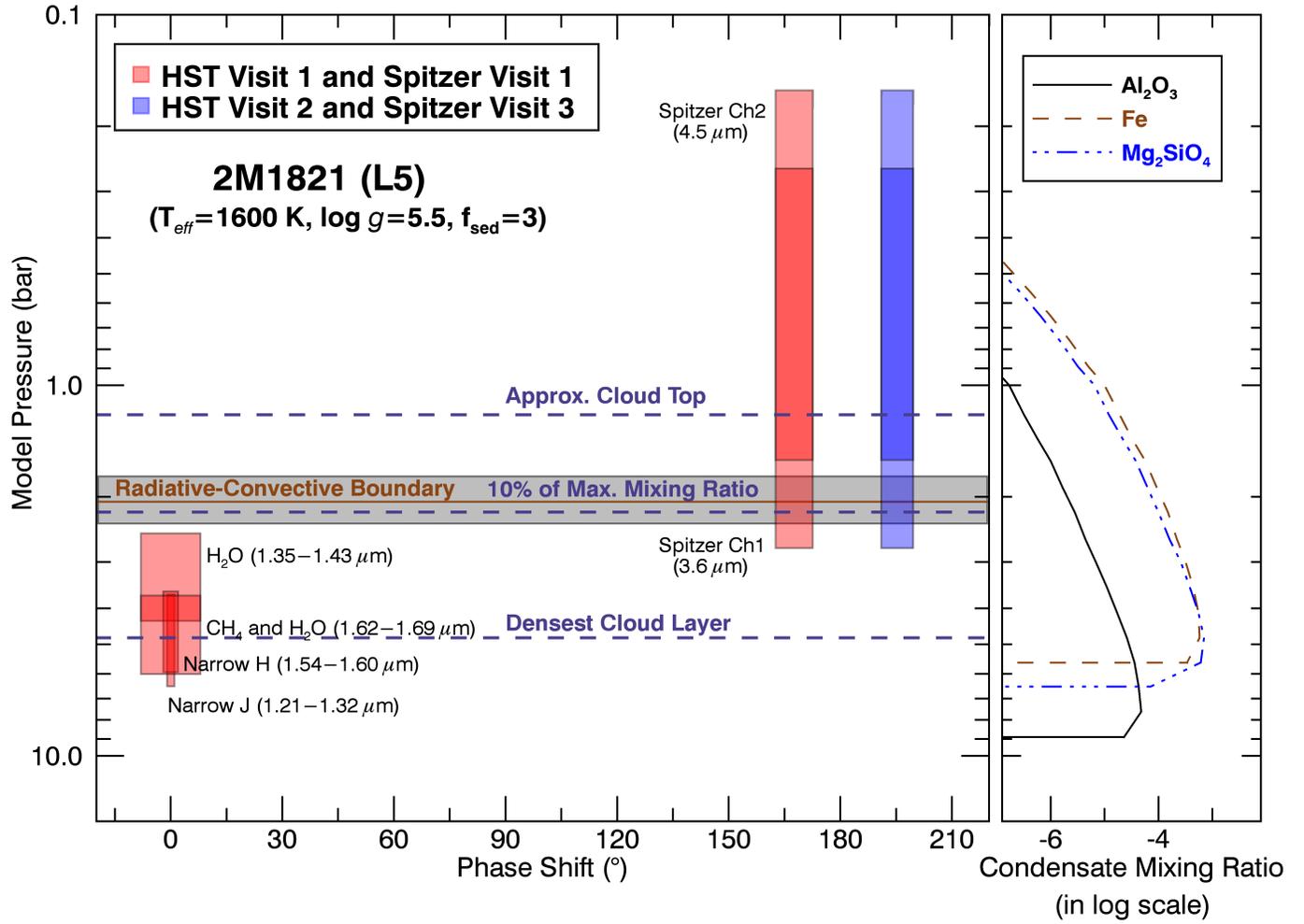}
       \caption{The same as Figure~\ref{shftpressure2228} except for 2M1821 (L5).}
           \label{shftpressure1821}
              \end{center}
    \end{figure*}

%%\begin{figure}[ht]
%%  \begin{center}
%%    \includegraphics[scale=0.65]{all_t1600g3000f3.ps}
%%   \caption{Relative flux contributions and cumulative contributions for the best-fit model of the L5 targets. 
%%                The model is for \Teff\ of 1600 K, \logg\ of 5.5, and $f_{\rm{sed}}$ of 3. See discussion in \S 6.1. 
%%               }
%%              \end{center}
%%    \end{figure}

\end{document}